\documentclass[sigconf]{acmart} 
\AtBeginDocument{%
  }

\usepackage[printonlyused,nolist,nohyperlinks]{acronym}
\usepackage{graphicx}
\usepackage{amsmath}
\makeatletter
\@namedef{ver@lstmisc.sty}{}
\makeatother

\usepackage{listings}
\usepackage{pifont}   
\usepackage{array}    
\usepackage{float}
\usepackage{multirow}
\usepackage{xcolor} 
\usepackage{url}

\copyrightyear{2026}
\acmYear{2026}
\setcopyright{cc}
\setcctype{by}
\acmConference[CHI '26]{Proceedings of the 2026 CHI Conference on Human Factors in Computing Systems}{April 13--17, 2026}{Barcelona, Spain}
\acmBooktitle{Proceedings of the 2026 CHI Conference on Human Factors in Computing Systems (CHI '26), April 13--17, 2026, Barcelona, Spain}
\acmDOI{10.1145/3772318.3791670}
\acmISBN{979-8-4007-2278-3/2026/04}

\begin{document}

\acrodef{AI}{Artificial Intelligence}
\acrodef{API}{Application Programming Interface}
\acrodef{HCI}{Human-Computer Interaction}
\acrodef{ANOVA}{Analysis of Variance}
\acrodef{LLM}{Large Language Model}

\title{From Use to Oversight: How Mental Models Influence User Behavior and Output in AI Writing Assistants}


\author{Shalaleh Rismani}
\authornote{Corresponding author: shalaleh.rismani@mail.mcgill.ca}
\affiliation{%
  \institution{McGill University}
  \city{Montr\'{e}al}
  \country{Canada}
}

\author{Su Lin Blodgett}
\affiliation{%
  \institution{Microsoft Research}
  \city{Montr\'{e}al}
  \country{Canada}}

\author{Q. Vera Liao}
\affiliation{%
  \institution{University of Michigan}
  \city{Ann Arbor}
  \state{Michigan}
  \country{USA}
}

\author{Alexandra Olteanu}
\affiliation{%
 \institution{Microsoft Research}
 \city{Montr\'{e}al}
 \country{Canada}}

\author{AJung Moon}
\affiliation{%
  \institution{McGill University}
  \city{Montr\'{e}al}
  \country{Canada}}

\renewcommand{\shortauthors}{Rismani et al.}

\begin{abstract}

AI-based writing assistants are ubiquitous, yet little is known about how users’ mental models shape their use. We examine two types of mental models---{\em functional} or related to {\em what} the system does, and {\em structural} or related to {\em how} the system works---and how they affect {\em control behavior}---how users request, accept, or edit AI suggestions as they write---and {\em writing outcomes}. 
We primed participants ($N=48$) with different system descriptions to induce these mental models before asking them to complete a cover letter writing task using a writing assistant that occasionally offered preconfigured ungrammatical suggestions to test whether the mental models affected participants' critical oversight. 
We find that while participants in the structural mental model condition demonstrate a better understanding of the system, this can have a backfiring effect: while these participants judged the system as more usable, they also produced letters with more grammatical errors, highlighting a complex relationship between system understanding, trust, and control in contexts that require user oversight of error-prone AI outputs.

\end{abstract}



\maketitle

\section{Introduction}

As \ac{AI} systems are increasingly integrated into everyday tasks, what must users understand about these systems in order to engage with them effectively and appropriately? Safe operation of complex, safety-critical technological systems (e.g., airplanes) typically demands their operators (e.g, pilots) to have a high degree of understanding of how a system works and how it can be used \cite{Silva2015-sw, Leveson2004-jn, Sarter2003-fv}. 
Prior work in human-computer interaction and system safety provides empirical evidence as to why users with more complete and accurate \textit{mental models}---their understanding of how a system works and how it can be used---are better equipped to monitor system behavior, anticipate failures, and intervene in near-incident scenarios \cite{Gaspar2020-xj,Silva2015-sw,Wiegand2019-tz}.

While these findings are well established in settings that have traditionally been deemed safety-critical (e.g., aviation), it remains unclear if, how, and when they translate to settings where generative AI systems are being used for everyday tasks (e.g., writing). 
Today, a variety of \ac{AI} systems (e.g., powered by \acp{LLM}) are integrated into writing platforms to provide writing support---for example, by generating next-sentence suggestions \cite{Lee2024-rb}---in order to improve efficiency and accessibility, especially for more inexperienced users~\cite{Li2024-jm,Goodman2022-jo,Gayed2022-fx}. 
In contrast to more traditional safety-critical systems, however, \ac{AI}-based writing assistants differ in key ways: 
their failures can be subtle or less salient; 
the consequences of such failures often do not lead to severe harm immediately \cite{Li2024-jm,Benharrak2024-yz,Jakesch2023-tx}; and developers often aim to make it frictionless and easier for users to write rather than encouraging agency and control \cite{Lee2024-rb,Reza2025-tt}.
This is not to imply that failures of these systems and their consequences in everyday writing tasks are inconsequential or that they should be accepted as the norm.
Although subtle, failures and consequences of \ac{AI}-assisted writing can include biased \cite{Jakesch2023-tx}, inaccurate or incoherent \cite{Sun2024-pl}, and 
low quality writing outputs \cite{Agarwal2025-mc,Li2024-yj}.
The use of such systems can also interfere with the writer's voice 
\cite{Draxler2024-ut,Kadoma2024-sv,robertson2021can}. These can all pose threats to a sense of authorship, ownership, and authenticity that writers continue to demand and value even as they adopt \ac{AI} tools \cite{robertson2021can,Kadoma2024-sv,Hwang2025-zo}.\looseness=-1 

Existing work on recommendation systems \cite{Kulesza2013-by} and predictive algorithms \cite{Bansal2019-cf} has shown that more accurate and thorough mental models of \ac{AI} systems can improve users’ satisfaction and trust, as well as support better decision making in task selection and prediction settings.
By extension, one can hypothesize that users with more expansive mental models of \ac{AI}-based writing assistants should be able to more effectively use and control these systems towards desired outcomes. Yet the role of such mental models in shaping interaction with LLM-based writing assistants remains underexplored, particularly in situations where outputs from such systems may be problematic, irrelevant, or erroneous.
To address this gap, we design and conduct a in-person between-subjects experiment ($N= 48$) to explore how users with different mental models of the same system---shaped by alternative descriptions---engage with \ac{AI}-based writing assistants. Our study and experiment design is guided by the following research questions:\looseness=-1
\begin{itemize}
    \item RQ1: How do different mental models affect the way users exert control in the writing process?
    \item RQ2: How do different mental models held by users influence the quality of the final written output?
    \item RQ3: How do these mental models shape users' overall experience of writing with the AI-based writing assistant?

\end{itemize}

To address these questions, we leveraged existing frameworks where mental models can be described as either \textit{functional} or \textit{structural} \cite{Kulesza2013-by}. While a functional mental model only involves an understanding of how to use the system, a structural mental model builds on this understanding and includes knowledge of how the system works.  
For our experiment, we use a modified version of the open-source CoAuthor platform \cite{Lee2022-hh}, which provides \ac{AI}-based next-sentence suggestions and supports meta-prompting. Prior to a cover letter writing task, we primed participants' mental models by showing them one of two videos: one on how to use the system (functional), and one also explaining how it works (structural). To better observe how participants exert control over the system, we intentionally introduced suggestions containing spelling or grammatical errors. We then analyzed participants' interactions with CoAuthor, final written work, and self-reported experiences from a post-task survey and semi-structured interview.
Within this staged erroneous output context, our results reveal a key tension between system understanding, trust, and oversight. Participants with structurally richer mental models reported higher perceived ease of use, but they produced letters with higher grammatical errors and tended to accept a higher rate of flawed suggestions. This pattern may reflect a form of overtrust, in which increased confidence in understanding the system results in users placing greater reliance on its suggestions than is warranted.

\noindent \textbf{Contributions.}
First, we present an empirical case study of how different mental models about an \ac{AI} writing assistant affect users' interactions with the system, particularly in the presence of low-quality suggestions. We further contribute a controlled experimental approach for manipulating and testing users’ functional and structural mental models through instructional priming under staged system failures.
Second, we provide a baseline empirical account showing that while a deeper system understanding may increase perceived ease of use and foster greater trust, it may also lead users to uncritically accept staged flawed suggestions that contain spelling and grammatical errors, resulting in lower writing quality, motivating the need for future studies to further examine connection between mental models, system failure, oversight, and trust in \ac{AI}-assisted writing.
Third, drawing from qualitative observations, we highlight the central role of interaction affordances in shaping perceptions of control and ownership as opposed to mental model understanding; users’ sense of agency may therefore be driven more by what systems allow them to do than by what they know about how those systems work.\looseness=-1

\section{Background}
This section reviews prior work on mental models, user control, and oversight from the perspectives of \ac{HCI} and system safety, as well as research on in AI-based writing assistants, to contextualize our study.  

\subsection{Mental Models and Their Role in Human–AI Interaction}

\textit{Mental models} have been a central concept in fields like \ac{HCI} \cite{Carroll1988-lz, Staggers1993-bn} and system safety \cite{Blokland2020-it, Cox2003-xi, Tarola2018-nm}. Mental models are internal representations that people form of \textit{target systems (e.g., computer)} when interacting with them. 
In cognitive psychology, mental models are understood to be partial, evolving, and sometimes inaccurate representations that people use to reason about complex systems under uncertainty \cite{Norman1987-yg, Johnson-Laird1983-ay}. Because these representations shape how users anticipate system states and interpret feedback, they directly influence trust, reliance, and decision making. \ac{HCI} and system safety literature has long adopted this concept to explain how people learn, use, and reason about interactive systems \cite{Rasmussen1987-aq, Carroll1988-lz}, emphasizing that mismatches between users’ mental models and a system’s actual behavior can hinder effective interaction and lead to erroneous outcomes, as discussed further in the next section.

Prior work shows that users’ mental models of AI systems are multifaceted and can reflect both functional expectations and deeper understanding about system behavior. People construct these models by combining explicit information, such as model explanations, with inferences drawn from their use, often producing simplified representations that support everyday reasoning. When examining users' mental models of recommendation systems, \citet{Kulesza2013-by} distinguish between \textit{functional mental models}, which represent how the system could be used, versus \textit{structural mental models}, which describe how the system works. \citet{Gero2020-dz} examine the mental models users form of \ac{AI} agents in a collaborative game and conclude that users' mental models of an \ac{AI} agent have three components: its behavior at a large scale, its knowledge of various topics, and its behavior at the scale of individual output. Recent research has examined users' mental models of \ac{LLM}- and generative \ac{AI}-based systems \cite{Wang2025MentalModelsGenAI,Mehmood2025-jh}. For example, \citet{Mehmood2025-jh} document heterogeneous mental models of \acp{LLM} at both individual and societal levels, ranging from optimistic views of \acp{LLM} as helpful tools to more skeptical views of them as suspicious actors.

Mental models influence how users interact with an \ac{AI} system and shape their experience and the final output. Research in \ac{HCI} and human–machine trust further shows that users’ expectations about a system’s capabilities and limitations strongly affect how much they trust and rely on it, with inaccurate or incomplete models leading to both over-reliance and unwarranted skepticism \cite{Gebru2022,Muir1987,Bach2024, Vodrahalli2022_DoHumansTrustAIMore}. Prior work on predictive algorithms has shown that persuasive explanations can lead users to form mental models that promote over-reliance or under-trust \cite{Nourani2021-jy, Bansal2019-cf, Bansal2021-cp, Bauer2023-sw}. For example, \citet{Bansal2019-cf, Bansal2021-cp} argue that explanations should prioritize informativeness over persuasion to help users develop calibrated trust and avoid inappropriate reliance.
Recent studies of \acp{LLM} show that when users hold incomplete or inaccurate expectations about how these systems behave, they may rely on outputs in ways that are not well aligned with task needs, such as accepting suggestions too readily \cite{Bo2025-bd,Kim2025-lp,Gerlich2025-ct}. 
Work on recommendation systems has further explored how strengthening a user's mental model can improve their experience; for example, \citet{Kulesza2013-by} found that providing structural scaffolding---offering insight into how a music recommendation system worked---helped users build more accurate mental models and led to higher user satisfaction with the system’s outputs. 
These works establish that mental models shape how users interact with \ac{AI} systems, the degree to which they trust and rely on these systems, and ultimately how these factors influence both their perceived experience and interaction outcomes. In the next section, we further elaborate on the connection between mental models and users’ ability to exercise control and oversight, focusing specifically on the case of \ac{AI}-based writing assistants. \looseness = -1

\subsection{User Control,  Oversight, and AI-based Writing Assistants}

As \ac{AI} systems become integrated into a wide range of applications, regulatory frameworks such as the EU \ac{AI} Act increasingly emphasize the need for effective human oversight: a user’s ability to notice, question, and intervene when system outputs may be inappropriate or harmful \cite{EUAIAct2024}. System safety research provides a longstanding foundation for understanding oversight, defining user control as the ability to intentionally influence, override, or intervene on system behavior to ensure that it operates safely and as intended \cite{Leveson2012-jp}. Inconsistent, inadequate, and incomplete user mental models of a system are recognized as key factors in hindering their ability to exert effective control \cite{Leveson2012-jp}. Decades of work in safety-critical domains---such as aviation, medical devices, and autonomous vehicles---demonstrate that how well a user understands a technological system impacts how they interact with and control it \cite{Dekker2017-gn, Dobbe2022-kf, Burtscher2012-om}. For example, drivers who understand the operational logic of adaptive cruise control are better able to override unsafe behaviors than those who only know how to operate its interface \cite{Gaspar2020-xj}. 
Given these illustrations of how inadequate and inconsistent mental models can lead to harm, system safety emphasizes the importance of designing environments and feedback mechanisms that help users recognize hazardous conditions---system states or interactions that increase the likelihood of harm---, monitor their actions, and recover prior to occurrence of an incident \cite{Leveson2012-jp, Dekker2017-gn}.\looseness=-1

Given the importance of effective oversight, together with extensive evidence that users’ mental models shape trust, reliance, outcomes, and user experience, it is critical to examine how mental models relate specifically to oversight of AI systems. In this work, we focus on this question in the context of \ac{AI}-based writing assistants, which are now widely used across domains ranging from creative writing to professional and scientific communication \cite{Yuan2022-gt, Ippolito2022-hx, Kim2023-pf, Gero2022-wr, Afrin2021-fk, Mirowski2023-qz, Goodman2022-jo, Guo2025-mn}.
In the domain of \ac{AI}-based writing assistants, these oversight challenges surface through cognitive and epistemic risks rather than physical ones. Specifically, \ac{AI}-generated suggestions can influence what users write and how they represent themselves. For instance, \citet{Poddar2023-ki} show that \ac{AI}-based writing assistants can shift self-presentation in personal bios, and \citet{Jakesch2023-tx} demonstrate that \ac{AI} suggestions can sway opinions in argumentative writing---even when misaligned with a user’s initial stance. Such influence may occur without users fully recognizing its extent, raising questions about users’ ability to oversee \ac{AI}-generated content and maintain ownership.  However, this line of work does not examine whether such influence varies as a function of users’ underlying mental models.

Complementary work in \ac{HCI} further shows that interface design can either support or undermine user control. Offering multiple parallel suggestions can aid idea generation but may also increase decision fatigue \cite{Buschek2021-xh}, while diegetic prompts can enable more intuitive interaction \cite{Dang2023-ug}. Although these studies highlight the intricacies of designing writing assistant interfaces that support effective control, they primarily focus on interaction mechanics or user preferences rather than on how users’ mental models influence their engagement with the system or how different control mechanisms shape those mental models.

Despite growing literature on influence, interaction patterns, and user experience in \ac{AI}-based writing assistants, prior work has not examined whether users with different mental models demonstrate different levels of control and oversight. Building on Draxler et al.’s definition of objective control---``the degree of influence that users have over the \ac{AI}-generated text, e.g., by employing interaction methods” \cite{Draxler2024-ut}---we interpret these interaction behaviors as concrete indicators of user control and oversight. This aligns with system safety perspectives, where user control refers to a user’s ability to actively influence or override system behavior to ensure task success and mitigate errors before harm occurs. 
Building on this framing, we examine how mental models shape users’ ability to exercise control and oversight in \ac{AI}-based writing assistants; the next two sections outline our study design.

\section{Study Overview and Hypotheses}
Drawing on prior work in system safety and \ac{HCI}, in this study we distinguish between \textit{functional} (focused on how to use a system) and \textit{structural} (focused on how the system works) mental models to design a controlled experiment \cite{Kulesza2013-by,Gero2020-dz}. Participants were asked to produce a high-quality cover letter under a time constraint using an \ac{AI}-based writing assistant. We chose this task because cover letters are a high-stakes, consequential form of writing, where errors, tone, and content quality can affect real-world outcomes such as job opportunities. The time pressure was introduced to encourage active engagement with the writing assistant. To better isolate the effects of mental model differences, we recruited participants with minimal prior experience using \ac{AI}-based writing assistants. To study how participants exercise oversight in a controlled and comparable way, we deliberately injected grammatical and spelling errors into the writing assistant’s suggestions. We chose this error type, rather than issues such as inappropriate tone or factual issues, because grammatical and spelling mistakes are relatively clear, easy to fix, and less subject to disagreement among participants. This allows us to see whether users actively scrutinize and refine \ac{AI}-generated suggestions rather than passively accept them. Further details about the experimental procedure, the recruitment process, the platform and the writing tasks are provided in Section \ref{method}. We present our research questions, expected outcomes, and corresponding hypotheses in Table \ref{tab:rq-hypotheses}.  \looseness=-1

\begin{table*}[!t]
\centering
\caption{Research questions, expected outcome, and hypotheses}
\def\arraystretch{0.9}
\setlength{\tabcolsep}{0.4em}
\footnotesize
\label{tab:rq-hypotheses}
\resizebox{\textwidth}{!}{
\begin{tabular}{@{}
  >{\raggedright}p{3.3cm}
  >{\raggedright}p{5.0cm}
  >{\raggedright\arraybackslash}p{9.2cm}
@{}}
\toprule
\textbf{Research Question} & \textbf{Expected outcome} & \textbf{Hypotheses} \\
\midrule
\textbf{RQ1: How do different mental models affect the way users exert control in the writing process?} & Participants with a structural mental model will demonstrate more effective and deliberate control, steering the system rather than passively accepting suggestions. & 
H1a: Structural participants will request more \ac{AI}-generated suggestions. \newline
H1b: Structural participants will accept a greater proportion of suggestions. \newline
H1c: Structural participants will make more edits to the accepted suggestions. \newline
H1d: Structural participants will issue more instructions to steer suggestions. \newline
H1e: Structural and functional participants will contribute an equal amount of original text. \\
\midrule
\textbf{RQ2: How do different mental models held by users influence the quality of the final written output?} & Participants with a structural mental model will produce higher-quality cover letters with fewer errors and better overall quality. & 
H2a: Structural participants will achieve higher grammar correctness scores and accept fewer erroneous suggestions. \newline
H2b: Structural participants’ letters will receive higher overall quality ratings.\\
\midrule
\textbf{RQ3: How do these mental models shape users’ overall experience of writing with the \ac{AI}-based writing assistant?} & Participants with a structural mental model will report more positive perceptions of the writing process and system, including greater perceived control, ownership, and satisfaction. & 
H3a: Structural participants will report higher perceived usefulness and ease of use. \newline
H3b: Structural participants will rate the quality of their final outputs more highly. \newline
H3c: Structural participants will report a stronger sense of control. \newline
H3d: Structural participants will report a greater sense of ownership. 
 \\
\bottomrule
\end{tabular}
}
\end{table*}

For control behavior \textbf{(RQ1)}, we expect that participants with a structural mental model will demonstrate more deliberate control over the system. Following Draxler et al.’s definition of control as the user’s influence over \ac{AI}-generated text \cite{Draxler2024-ut}, this behavior can manifest through (1) requesting, accepting, or editing \ac{AI} suggestions in the text editor and (2) providing explicit instructions in the Instruction Box (refer to Section \ref{coauthordescription} for a description of the platform). Because structural participants understand how the system works and what it can and cannot do, we expect them --- especially under time pressure to complete the task --- to request more suggestions, steer those suggestions through targeted instructions, accept more of them, and then make edits to refine the output to their exact preferences, thereby ensuring higher quality in the final cover letter. We do not expect the overall proportion of text selected from \ac{AI}-generated suggestions versus written by participants differ across both conditions. 
For writing quality \textbf{(RQ2)}, we hypothesize that participants with a structural mental model will produce higher-quality cover letters. Prior work suggests that users who better understand system functioning can more effectively guide outputs \cite{Kulesza2013-by}. We therefore expect structural participants to exercise stronger oversight, accept fewer erroneous suggestions, and ultimately produce cover letters with fewer errors and higher overall quality.
For perceptions and experience \textbf{(RQ3)}, we anticipate that structural participants will report more positive evaluations, including greater confidence in the final output, a stronger sense of control and ownership, and higher ratings of ease of use. All the corresponding constructs, data sources, and measures are outlined in Section \ref{sec:data-analysis} and Table \ref{tab:hypotheses-measures}. \looseness=-5

\section{Study Design and Methodology}
\label{method}
We conducted an in-person controlled experiment where we asked participants ($N = 48$) to write a cover letter using an \ac{AI}-based writing assistant, a modified version of the CoAuthor platform \cite{Lee2022-hh}. Participants were randomly assigned to view one of two video descriptions of the assistant, which were designed to elicit distinct mental models: (1) a {\em functional} description that focused solely on {\em how to use the platform}, and (2) a {\em structural} description that additionally explained {\em how the platform works}. For this, we deliberately recruited individuals who self-reported limited familiarity with \ac{AI}-based writing assistants, allowing us isolate the effects of our experimental manipulation. To evaluate participants’ ability to exercise critical oversight, we intentionally embedded erroneous suggestions (e.g., grammar and spelling mistakes) into the \ac{AI}-based writing assistant’s suggestions at a fixed frequency. We compared participants' behavioral data, cover letter quality, and self-reported experiences across the two mental model conditions. All participants provided informed consent, and the study was approved by the Research Ethics Board of McGill University (REB \#23-08-063; \textit{Mental Models and User Control When Interacting with \ac{AI} Systems}).

\begin{figure*}[!t]
    \centering
    \includegraphics[width=1\textwidth]{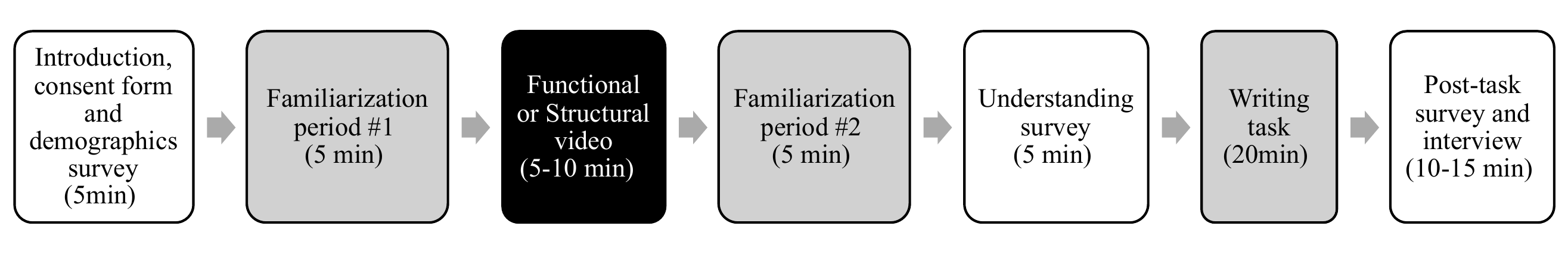}
    \caption{This figure illustrates the experimental design and procedure. White boxes indicate stages where participants completed various surveys. Gray boxes represent phases where participants either familiarized themselves with the platform or engaged in the writing task. The black box marks the manipulation point in the experiment.}
    \Description{Flowchart of the experimental procedure. Participants first completed the consent form and the demographics survey (5 min), then familiarization period \#1 (5 min), followed by either a functional or structural video (5–10 min, manipulation point). Next came familiarization period \#2 (5 min), an understanding survey (5 min), the writing task (20 min), and finally a post-task survey and interview (10–15 min). White boxes indicate surveys, gray boxes indicate familiarization or writing, and the black box marks the manipulation.}
    \label{fig:ch6-hai-mm-study}
\end{figure*}
\subsection{Participant Recruitment}

Participants were recruited through posters, organizational mailing lists, and dedicated social media groups where individuals had previously expressed interest in research studies. To inform our sample size, we conducted an in-person pilot study (\textit{n =} 8) and calculated the ratio of accepted suggestions that contained grammatical errors to the total number of accepted suggestions. We observed a notable difference between the mental model conditions: for participants in the functional condition, 22\% of accepted suggestions contained grammatical errors, versus 11\% in the structural condition (effect size = 0.83). Based on this effect size, and assuming $\alpha = .05$ and power = .80, our power analysis indicated a minimum of 19 participants per condition (38 total) for the main study. 

We recruited participants with prior experience with writing cover letters and regular use of English in academic or professional settings.
We also selectively recruited individuals who reported limited understanding of \ac{AI}-based writing assistants, recognizing that experienced users often carry pre-existing mental models of such systems. By working with participants who had minimal prior understanding, we ensured that our experimental manipulation (two different system descriptions) could shape their mental models, allowing us to study how these influenced engagement with the system. 
Eligibility was determined through four screening questions: (1) experience writing cover letters (Yes/No); (2) daily use of English in professional or academic setting (Yes/No); (3) familiarity with \ac{AI}-based writing assistants (Not at all familiar, Slightly familiar, Moderately familiar, Extremely familiar), and (4)  level of understanding of how \ac{AI}-based writing assistants work (Very poor, Poor, Fair, Very good, Excellent). Participants who answered “yes” to Questions 1–2, “not at all familiar” or “slightly familiar” for Question 3, and “very poor,” “poor,” or “fair” for Question 4 were eligible. Of the total of 295 individuals who completed the intake form, 72 met these criteria. From this pool, we scheduled 48 participants -- 24 per mental model condition -- for the in-person study, allowing for missing and outlier data points.
 
 Across the two conditions, the participants were demographically comparable.  The average age was 24.7 years ($M = 24.7$, $SD = 8.3$) in the functional condition and 23.9 years ($M = 23.9$, $SD = 3.1$) in the structural condition. Two-thirds of the participants identified as female (66.7\%), 22.9\% as male, 6.3\% as non-binary, and 4.2\% preferred not to answer, with nearly identical distributions between conditions. Just over half of the participants (54.2\%) reported English to be their native language. The functional condition had a higher proportion of native English speakers (66.7\%) compared to the structural condition (41.7\%). However, a chi-square test of independence indicated that this difference is not statistically significant, $\chi^2(1, N = 48) = 3.021$, $p = .082$. Although this result falls short of conventional significance thresholds, the imbalance in native English speakers across conditions may still constitute a potential limitation. We note this here and considered it in subsequent analyses.
Participants reported writing in English frequently ($M = 4.67$, $SD = 0.56$, on a 5-point scale) and expressed high confidence in their English writing ability ($M = 4.50$, $SD = 0.65$). Self-assessed cover letter writing ability was rated between ``fair'' and ``very good'' on average ($M = 3.60$, $SD = 0.68$), and confidence in writing cover letters was rated as just above ``fair'' ($M = 3.21$, $SD = 0.65$). As expected, the use of \ac{AI}-based writing tools was generally infrequent ($M = 2.48$, $SD = 1.17$), with no notable differences between conditions. 

\subsection{Experimental Design}
Our experiment followed a between-subjects design with mental model (functional vs. structural) as the independent variable (see Figure \ref{fig:ch6-hai-mm-study}). We operationalized this variable by priming participants with two different video descriptions about the modified CoAuthor platform, one intended to prime a functional and the other a structural mental model.

The study was conducted in person to ensure that participants engaged fully with the writing task without relying on external \ac{AI}-based tools and to maintain experimental control by avoiding well-documented pitfalls of online \ac{HCI} studies such as reduced attention, multitasking, unmonitored tool use, and data-quality concerns in crowdsourced settings \cite{Kittur2008-yf,Panicker2024-jo, Paolacci2014-st}. Participants first completed a survey capturing demographics, self-assessed English writing skills, and prior experience with cover letters and \ac{AI}-based writing assistants. These data were collected to inform and contextualize our findings. For an initial familiarization, the experimenter briefly introduced the modified CoAuthor platform and guided the participants through the main functionalities such as writing in the editor and pressing \textit{tab} to get suggestions. Next, participants were randomly assigned to watch one of two pre-recorded videos---one designed to elicit a functional mental model (how to use the system), and the other a structural mental model (how the system works), as described in Section \ref{mm_manipulation}. The functional video was 3:58 minutes, and the structural video was 8:54 minutes; although the videos differed in length, this reflected the additional explanation required to convey how the system works. Because our goal was to examine whether participants who received a more detailed explanation of the system’s underlying mechanisms would form different mental models and engage with the writing assistants differently, we accepted this difference in duration as appropriate for the manipulation.
Afterward, participants explored the platform independently and could ask questions during a second familiarization period.
Finally, before beginning the writing task, participants completed a brief survey to assess their understanding of the platform, which served as a manipulation check.

At this point, participants in both conditions were presented with a job posting and had 20 minutes to write a cover letter using the modified CoAuthor platform, described in 
Sections \ref{cover-letter-rubric} and \ref{coauthordescription}. The allotted writing time was kept the same regardless of the length of the pre-recorded video across the two conditions, and participants in the structural condition were not under additional time pressure to complete the writing task. During the task, the platform injected grammatical and spelling errors at a predetermined frequency, allowing us to observe how participants with different mental models detected and addressed such mistakes. After the writing task, the participants completed a post-task survey to report on their writing experience and perceptions of the platform and final output. This post-task survey was adapted from prior validated studies on writing assistants that examine perceived writing quality, user experience, sense of control, and ownership \cite{Draxler2024-ut,Lee2022-hh,Kadoma2024-sv,Dang2023-ug,Buschek2021-xh}.
Lastly, the experimenter conducted a semi-structured interview with the participants to gather richer, qualitative insights into their experience of using the platform and writing the cover letter.  The questions from the three surveys are included in Appendix \ref{app:demographics}, \ref{app:understanding}, and \ref{app:posttask}. The experimenter script, including the post-task interview questions, the mock job posting, the pre-recorded video slides, and the videos are all included in the supplemental material.\looseness=-1

\begin{figure*}[!t]
    \centering
    \includegraphics[width=0.75\textwidth]{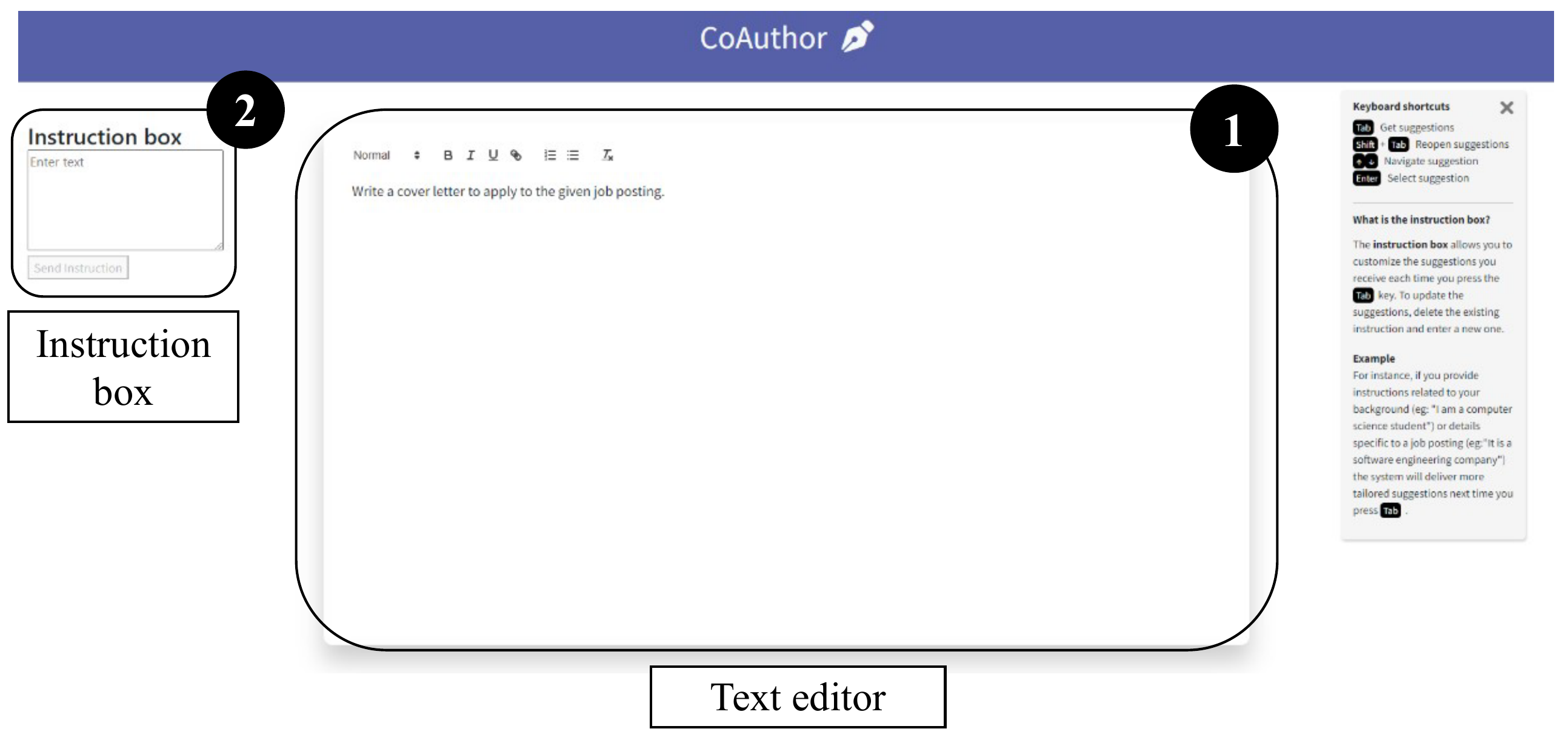}
    \caption{A screenshot of the AI-based writing assistant with the text editor where the main letter is written in the center, the Instruction Box is on the left side, and a basic summary of how to use the platform is on the right side. The interface is a modified version of the open-source CoAuthor platform created by \citet{Lee2022-hh}.}
    \Description{Screenshot of the modified CoAuthor platform. A large text editor occupies the center where participants write the cover letter. On the left is a small instruction box where participants can write meta-prompts to modify the suggestions that appear in the text editor when they press the tab button. On the right, a sidebar gives a basic summary of how to use the platform.}
    \label{fig:ch8-coauthor-pic}
\end{figure*}

\subsection{The Cover Letter Writing Task}
\label{cover-letter-rubric}
Participants were asked to write a high-quality cover letter for a Customer Service Representative position at a fictional company, Radiant Solutions---a technology solutions provider across various sectors. The cover letter writing task was chosen because it has a clear functional objective (to secure an interview) and audience (recruiter/potential employer). We asked participants to write a letter that illustrates their interest in the role and helps them secure an interview.
Since all participants had prior experience writing cover letters, we were not prescriptive about what participants should consider as a high-quality letter. We instructed participants to write about themselves (rather than a fictional character), include at least one personal experience, and aim to fill the full span of the text editor, encouraging a letter of approximately 300 words. They were given 15 minutes to complete the task with the option to request an additional 5 minutes if needed. Lastly, the job posting was intentionally designed to be broad and accessible, enabling participants with diverse backgrounds, skill sets, and educational experiences to respond meaningfully. We initially generated the job posting using GPT-4o by prompting it to ``create a generic job posting for a customer sales representative,'' and then refined the output to align with the style and content of job postings returned when searching for \textit{customer sales representative} positions on LinkedIn and Indeed.\looseness=-1

\subsection{Platform Description}
\label{coauthordescription}

We adapted and extended the open-source CoAuthor platform \cite{Lee2022-hh} to create the writing assistant used in this study. In the modified platform, participants typed freely in the main text editor and pressed the \textit{Tab} key whenever they wanted the system to propose continuations. The interface then displayed up to five candidate suggestions, and participants could choose to insert one into their draft or ignore them and continue writing. Participants could also optionally provide guidance through an Instruction Box that allowed them to customize the generated suggestions.

The modified platform preserved CoAuthor’s core mechanism for generating suggestions. Each \textit{Tab} press triggered an \ac{API} call to a large language model, sending up to 4{,}000 tokens of recent editor text as a prompt. The model returned 15 candidate suggestions. We used the GPT-3.5-turbo-instruct model, which OpenAI recommended at the time of the study for sentence-level completions. After filtering duplicates and blocked content, five suggestions were randomly selected and displayed to the user (Figure~\ref{fig:ch8-coauthor-pic}).

The modified platform was created by introducing two changes to the original CoAuthor system: (1) the controlled injection of suggestions containing grammatical and spelling errors at a fixed ratio, and (2) the addition of an \textit{Instruction Box}, which provides a meta-prompting functionality allowing users to steer the generated suggestions. We describe these two modifications next.

\subsubsection{Modification 1: Grammatical and spelling errors in the suggestions}

In this modification, CoAuthor was adapted to inject grammatical and spelling errors into \textit{two of every five suggestions} shown to participants. To do so, we introduced a second \ac{API} call, prompting the model to rewrite two suggestions with one grammatical or spelling error using a custom-designed prompt. The two suggestions were selected at random from the list of five suggestions. To ensure that the rewritten suggestions consistently contained a single deliberate error, we crafted prompts and piloted them to evaluate the number and type of grammatical or spelling errors they produced. We iteratively refined the prompt wording to ensure the modified CoAuthor platform reliably introduced exactly one error in each of the two regenerated suggestions. With \texttt{grammar\_mistakes} set to be 1, the final prompt we used was: 
\begin{lstlisting}
Rewrite "\"" + suggestion + "\"" with exactly " + str(grammar_mistakes) + " grammatical or spelling mistakes, including a malapropism or a subject-verb agreement error within the first 8 words."
\end{lstlisting}

\subsubsection{Modification 2: The Instruction Box}

For the Instruction Box, the user's input was concatenated with the text editor's content before being sent to the model to guide suggestion generation. Specifically, we added logic such that if a user provided an instruction, CoAuthor constructed a new prompt by prepending the instruction to the text in the editor (i.e., the prompt) using the following structure: \begin{lstlisting}
Consider " + UserInput + "\nGive suggestions for the prompt: " + prompt
\end{lstlisting} 

The reformulated prompt was then passed to the \ac{API} to generate completions as described earlier. We tested several variations and found that simple phrasing (e.g., “Consider”) was sufficient for the model to incorporate user instructions without altering their meaning. If no user instruction was provided, the platform defaulted to the editor text. 

With these modifications, participants could interact with CoAuthor by asking for suggestions, accepting or ignoring suggestions, adding or deleting content directly in the text editor, and using the Instruction Box to adjust the suggestions.\looseness=-3

\subsection{Mental Model Manipulation and Validation}
\label{mm_manipulation}

\begin{table*}[!t]
\centering
\caption{The content of the video description for the two experimental conditions}
\vspace{-6pt}
\footnotesize
\begin{tabular}{p{9cm}|c|c}
\textbf{Content} & \textbf{Functional} & \textbf{Structural} \\
\hline
What is the CoAuthor platform? & \ding{51} & \ding{51} \\
\hline
What are the key features of the CoAuthor platform? & \ding{51} & \ding{51} \\
\hline
What are the main functionalities of the CoAuthor platform?  & \ding{51} & \ding{51} \\ \hline
Usage of text editor (feature 1)& \ding{51} & \ding{51} \\
\hline
How does CoAuthor generate its suggestions?  & \ding{55} & \ding{51} \\
\hline
Examples of how content in the text editor affects the suggestions generated  & \ding{55} & \ding{51} \\
\hline
Usage of Instruction Box (feature 2) & \ding{51} & \ding{51} \\
\hline
How does the instruction from the Instruction Box customize the suggestions? & \ding{55} & \ding{51} \\
\hline
Examples of how content in the Instruction Box affects the suggestions generated  & \ding{55} & \ding{51} \\
\hline
\end{tabular}
\label{tab:ch6-coauthor-video-desc}
\end{table*}

\begin{table*}[!t]
\centering
\def\arraystretch{0.9}
\setlength{\tabcolsep}{0.4em}
\caption{The seven questions in the manipulation survey and their corresponding correct answers. Questions 5--7 assessed knowledge of internal system mechanisms described only in the structural video.}
\label{table:manipulation-check}
\vspace{-6pt}
\label{tab:manipulation-check}
\footnotesize
\begin{tabular}{@{}p{0.2cm}|p{9cm}|p{2cm}@{}}
\textbf{\#} & \textbf{Question} & \textbf{Correct Answer} \\
\hline
1 & CoAuthor is a writing assistant that does not rely on \ac{AI}. & False \\
\hline
2 & CoAuthor offers phrases and sentences when a user requests suggestions. & True \\
\hline
3 & Where should the user write the main body of their cover letter on the CoAuthor platform? & Text editor \\
\hline
4 & CoAuthor will automatically generate suggestions for the user when the user stops typing. & False \\
\hline
5 & CoAuthor filters and trims the original list of suggestions generated by the large language model before showing it to the user. & True \\
\hline
6 & CoAuthor gives suggestions based on what is a likely continuation of the last sentence in the text editor. & True \\
\hline
7 & The content of the Instruction Box is sent to a different large language model than the text in the text editor. & False \\
\hline
\end{tabular}
\end{table*}
To elicit the two mental models (functional and structural), we created a video for each condition, consisting of slides with a voiceover that provided descriptions and illustrated examples of the modified CoAuthor platform. The descriptions were developed iteratively, refined through pilot studies, and informed by key principles for supporting mental model formation in \ac{AI} systems, as described by Kulesza et al. \cite{Kulesza2013-by} and Gero and Möller \cite{Gero2020-dz}. 

The functional description video provided an overview of the functionalities of the platform, such as how the user can ask for suggestions, how they can customize suggestions using the Instruction Box, and other basic usage information. The structural description video included all these slides but added more details about the inner workings of the platform. For example, every participant was told that CoAuthor generates suggestions when pressing \textit{Tab} in the text editor. However, the structural description also covered that a \ac{LLM} is used to generate a set of original suggestions, which are filtered before being shown to the user. The structural video also provided examples to illustrate how the platform works. For instance, it included examples of how different starter sentences in a cover letter would lead to varying suggestions based on the content of the starter sentences. To account for the fact that the participants might have a pre-existing mental model of commonly used \ac{AI}-based writing assistants, both functional and structural descriptions highlighted how the modified CoAuthor platform used in the experiment differs from popular and commercially available tools such as ChatGPT and Grammarly. Table \ref{tab:ch6-coauthor-video-desc} illustrates the content that was and was not covered in the functional and structural video descriptions.\looseness=-1

To validate the effect of our manipulation, participants were asked to fill out an understanding survey, which served as a manipulation check (see Figure \ref{fig:ch6-hai-mm-study}), assessing participants’ understanding of the platform and the information presented in the instructional videos. As shown in Table~\ref{tab:manipulation-check}, Questions 1--4 asked about general usage and functionalities of the platform. These aspects were covered in both video descriptions, and therefore participants in both the functional and structural groups were expected to answer these items correctly. Questions 5--7 served as the manipulation check: they probed at how the system works as explained only in the structural video. As a result, participants in the structural condition were expected to answer these items correctly, whereas participants in the functional condition were expected to show lower accuracy, which could appear as either incorrect responses or selecting ``I don't know.'' For all questions except Question 3, participants selected ``True,'' ``False,'' or ``I don't know.'' For Question 3, they chose between ``Text Editor,'' ``Instruction Box,'' ``Both,'' or ``I don't know.''\looseness=-1

\subsection{Data Analysis}
\label{sec:data-analysis}

Guided by our research questions, we analyzed both quantitative and qualitative data collected from participants’ interactions with the writing assistant, their final cover letters, post-task surveys, and semi-structured interviews. Next, we describe the key measures, and the analysis approaches.

\subsubsection{Quantitative Analysis}
\label{ss:quant}

Our quantitative analysis was structured around the three research questions, each paired with specific hypotheses and constructs. 
Table~\ref{tab:hypotheses-measures} provides an overview of how each hypothesis maps to key constructs, corresponding data sources, and specific measures.\looseness=-1

For RQ1, we expect participants with a structural mental model to exert more deliberate control by requesting, steering, accepting, and editing more suggestions, while also contributing original text. We study this research question through the following measures and hypotheses:

\begin{itemize}
    \item \textbf{H1a (Requesting suggestions):} Measured by the total number of requested suggestions divided by the word count across the full document (\textit{suggestion requests}); participants in the structural condition will request more \ac{AI}-generated suggestions than those in the functional condition.

    \item \textbf{H1b (Accepting suggestions):} Measured by the total number of accepted suggestions over the total number of requested suggestions across the full document (\textit{acceptance ratio}); participants in the structural condition will accept a greater proportion of suggestions.

    \item \textbf{H1c (Editing accepted suggestions):} Measured by summing the number of deleted characters within the first 20 recorded edit actions after each suggestion is accepted (across the full document) and dividing by the total number of characters in all accepted suggestions (\textit{edit ratio}); participants in the structural condition will make more direct edits to accepted suggestions.

    \item \textbf{H1d (Steering via instructions):} Measured by the total number of instructions issued divided by the word count (\textit{instruction count}) in the writing session; participants in the structural condition will issue more instructions to steer \ac{AI}-generated suggestions.

    \item \textbf{H1e (Direct writing contribution):} Measured by the total number of user-written characters divided by the total number of characters in the final document (\textit{user character ratio}); participants in the structural and functional condition will contribute an equal level of original text.
\end{itemize}

For RQ2, we expect participants with a structural mental model to write higher-quality cover letters. We evaluated the letters using two complementary methods. 

\begin{itemize}
    \item \textbf{H2a (Grammar correctness):}  Measured by the Grammarly correctness score (normalized by total word count), which captures the number of suggested corrections related to grammar, spelling, and punctuation (\textit{Calibrated Corrections}) \cite{grammarly}. Participants in the structural condition are expected to require fewer corrections in their final letter. As a complementary exploratory analysis, we also examine the \textit{error ratio}, measured as the total number of accepted erroneous suggestions divided by total number of accepted suggestions across the full document, and \textit{errors per word}, measured as the total number of accepted suggestions containing grammatical errors divided by the word count of the full document.\looseness=-1
    \item \textbf{H2b (Overall quality):} Measured by two researchers (both co-authors of this paper) independently assessing writing quality using a structured rubric with three dimensions: \textit{relevance} of the skills and qualifications to the job posting, \textit{flow and clarity} of ideas, and \textit{tone/style appropriateness}. One researcher rated all of 48 letters, and a second rated 50\% for reliability. Inter-rater agreement was high, with weighted Cohen’s Kappa scores of 0.79 for \textit{relevance}, 0.70 for \textit{flow}, and 0.85 for \textit{tone/style}. The rubric for this qualitative quality assessment was grounded in participants’ own definitions of high-quality writing, including clearly aligning qualifications with the job, using a persuasive and sincere tone, and maintaining a concise, well-structured format. Participants in the structural condition are expected to produce letters that score higher on these dimensions. The full rubric description is provided in Appendix \ref{app:rubrics}.  \looseness=-1

\end{itemize}

For RQ3, we expect participants with a structural mental model to report more positive perceptions of the final outcome and the writing process. All measures for RQ3 were captured through 5-point Likert-scale items in the post-task survey. The survey was developed based on questions used in relevant prior work \cite{Draxler2024-ut,Kadoma2024-sv,Buschek2021-xh,Dang2023-ug,Lee2022-hh}.

\begin{itemize}
    \item \textbf{H3a (Perceived usefulness and ease of use):} Measured by survey items on usefulness and ease of use; participants in the structural condition will report higher ratings on both.
    \item \textbf{H3b (Perceived quality of the final output):} Measured by survey items on satisfaction with the final letter; participants in the structural condition will rate their letters more highly.
    \item \textbf{H3c (Perceived control):} Measured by survey items on perceived influence over the writing process and system suggestions; participants in the structural condition will report a stronger sense of control.
    \item \textbf{H3d (Perceived ownership):} Measured by survey items on authorship and attribution; participants in the structural condition will report higher ownership, indicating the letter reflects their own voice and intent. \looseness=-1

\end{itemize}

\begin{table*}[!t]
\centering
\footnotesize
\def\arraystretch{0.9}
\setlength{\tabcolsep}{0.4em}
\caption{Mapping of hypotheses to constructs and measures (quantitative analysis)}
\vspace{-6pt}
\label{tab:hypotheses-measures}
\resizebox{\textwidth}{!}{
\begin{tabular}{@{}
  >{\raggedright}p{1.2cm} 
  >{\raggedright}p{3.6cm} 
  >{\raggedright}p{2.7cm} 
  >{\raggedright\arraybackslash}p{9.3cm}
@{}}
\toprule
\textbf{Hypothesis} & \textbf{Construct(s)} & \textbf{Data Source} & \textbf{Specific Measures (calculated per participant and letter)} \\
\midrule
\textbf{H1a} & Requesting suggestions & Interaction logs & \textit{Suggestion requests:} total number of requests for suggestions divided by word count. \\
\textbf{H1b} & Accepting suggestions & Interaction logs & \textit{Acceptance ratio:} accepted suggestions divided by total suggestion requests. \\
\textbf{H1c} & Editing accepted suggestions & Interaction logs & \textit{Edit ratio:} number of deleted characters within the first 20 recorded edit actions after each accepted suggestion, divided by the total number of characters in all accepted suggestions. \\
\textbf{H1d} & Steering via instructions & Interaction logs & \textit{Instruction count:} total number of instructions the user inputs divided by the word count. \\
\textbf{H1e} & Direct writing contribution & Interaction logs & \textit{User character ratio:} total number of user-written characters divided by total characters.\\
\midrule
\textbf{H2a} & Grammatical correctness of the letter and accepted erroneous suggestions& Grammarly scores of the final letter, Interaction logs & \textit{Calibrated corrections:} Grammarly correctness score, defined as the number of corrections suggested by Grammarly divided by word count; \textit{Error ratio:} total number of accepted suggestions with errors divided by total accepted suggestions; \textit{Errors per word:} total number of accepted suggestions containing grammatical errors divided by word count.  \\
\textbf{H2b} & Overall cover letter quality & Qualitative quality assessment& \textit{Rubric-based quality ratings:}  a score between 1(poor) -- 5(excellent) along three dimensions of relevance, flow, and tone/style. \\
\midrule
\textbf{H3a} & Perceived usefulness and ease of use & Post-task survey & \textit{Usefulness and ease-of-use:} perceived helpfulness of the system and effort required to use it, assessed using multiple items rated on 5-point Likert scales.\\
\textbf{H3b} & Perceived quality of final output & Post-task survey & \textit{Perceived quality:} satisfaction with the quality of the final letter, assessed using multiple items rated on 5-point Likert scales. \\
\textbf{H3c} & Perceived control & Post-task survey & \textit{Perceived control:} sense of influence over the writing process, assessed using multiple items rated on 5-point Likert scales. \\
\textbf{H3d} & Perceived ownership & Post-task survey & \textit{Perceived ownership:} extent to which the letter reflects the participant’s own voice and intent, assessed using multiple items rated on 5-point Likert scales. \\

\bottomrule
\end{tabular}
}
\vspace{-10pt}
\end{table*}

\subsubsection{Qualitative Analysis}

To enrich and contextualize our quantitative findings, we conducted two qualitative analyses: one focused on participants’ written instructions to customize suggestions on the platform, and the other on the responses from the post-task semi-structured interviews.

First, to inform RQ1, we analyzed a total of 315 instructions written by participants during the cover letter revision task. To characterize how participants attempted to guide the AI-generated suggestions, we developed a coding rubric with three binary categories: whether the instruction aimed to shift the tone or style of the letter, articulate relevant qualifications, or prompt a structural change. Three researchers, who are all authors of this paper, collaboratively developed the rubric by jointly coding an initial set of 20 instructions. Two of the researchers then independently co-coded 40 instructions to establish consistency before the lead researcher completed coding the remainder. Inter-rater reliability was high across all dimensions ($\kappa = 0.775$ for \textit{tone/style shift}, $0.773$ for \textit{qualifications articulation}, and $0.713$ for\textit{ structure prompting}). This analysis provided insights into the type of instructions the participants used to steer the \ac{AI}-generated suggestions. The full rubric description is provided in Appendix \ref{app:rubrics}. 

Second, to inform all three RQs, we conducted a reflexive thematic analysis of the semi-structured interviews \cite{Braun_Clarke_2006, Braun2023-vs}, which explored participants’ perceptions of ownership, control, and writing quality. The coding process was explicitly oriented around themes relevant to our research questions. The lead researcher coded all interview transcripts, while a second researcher coded a quarter of them to support synthesis. Through iterative discussion, the two researchers compared interpretations, resolved discrepancies, and refined the coding structure until a shared understanding was reached. This process resulted in a set of themes describing how participants understood the role of \ac{AI}-based writing assistants, the degree of influence they felt over the evolving text, and the aspects of system interaction that shaped their sense of control and ownership.\looseness=-1

\section{Findings}
We begin by describing participants’ baseline writing practices and attitudes toward \ac{AI}-based writing assistants, followed by the validation results from our manipulation check to confirm the effectiveness of the mental model priming. We then present findings from both our quantitative and qualitative analyses, organized around the three research questions, and conclude with a reflection on study limitations and the researchers’ positionalities to contextualize our interpretations.\looseness=-1

\paragraph{Participants’ writing practices and perspectives on \ac{AI} assistance}

Participants in both mental-model conditions stated that they typically begin by brainstorming and outlining, followed by iterative refinement, when completing writing tasks relevant to their professional and personal contexts. Many reported using comparable strategies for cover letters, though several emphasized relying on templates or examples, noting that they often “start with the same two sentences” (R2-S) \footnote{Participant IDs ending in –F correspond to the functional condition, while those ending in –S correspond to the structural condition.} or “look for an example” (R2-S, R3-F). 

Most participants reported rarely using \ac{AI}-based writing assistants. Among participants who occasionally use these assistants, Grammarly and ChatGPT were most commonly mentioned. Notably, some acknowledged that \ac{AI}-based assistants such as Grammarly are \textit{``always on"} (R1-F) and present \textit{``in the background"} (R29-F), whether they actively sought them or not. Across both conditions, participants' overall attitudes were cautious or negative towards the use of \ac{AI}-based writing assistants. Concerns included loss of authenticity (e.g., \textit{``when I read it back, it doesn't feel authentic to me"} R37-F), ethical implications (e.g.,\textit{``I feel like a fraud,''} R3-F), privacy risks (e.g.,\textit{``I have to be careful with where I store information,"} R27-F), and the reliability of \ac{AI}-generated content, especially in higher-stakes writing such as \textit{``academic [assignments]"} (R1-F). Some participants mentioned \ac{AI}’s usefulness for low-stakes or templated tasks. Among all participants, the most common reasons for using \ac{AI}-based assistants were grammar and spelling checks, as well as phrasing improvements. 
A few participants also described using \ac{AI} for tasks like brainstorming ideas or adjusting tone, though these uses were less common.

All participants valued maintaining control over their writing. Many did not seek external help and avoided using \ac{AI}-based assistants to preserve this control, while some welcomed support from peers, templates, or \ac{AI}-based assistants for phrasing and high-level structure. Across both conditions, participants emphasized ensuring that their own voice remained clear and that they guided the ideas and overall direction. This preference extended to cover letter writing, though a few noted feeling less control “because they based [the letter] off of their job description” (R10-S). 
\looseness =-1

\paragraph{We observed a significant manipulation effect based on the manipulation check survey results} 
A Mann–Whitney U test indicates that participants in the structural condition (\( \text{Mdn} = 6.0 \), \( \text{Mean Rank} = 30.56 \)) outperformed those in the functional condition (\( \text{Mdn} = 5.0 \), \( \text{Mean Rank} = 18.44 \)) on the three manipulation check questions
(\( U = 142.50 \), \( Z = -3.15 \), \( p < .01 \)). 
This helped validate that our video-based manipulation influenced our participants' mental models.
At the same time, the structural group performed more poorly than the functional group on Q1 (i.e., ``CoAuthor is a writing assistant that does not rely on AI,'' see Table~\ref{table:manipulation-check}), which assessed a basic factual claim about CoAuthor (\( U = 228.00 \), \( Z = -2.34 \), \( p < .05 \)). Although unexpected, this single-item difference is unlikely to indicate a true misunderstanding of the platform. Post-task interviews confirmed that all participants recognized CoAuthor as an AI-based system. We conjecture that Q1 was likely misread or overlooked (e.g., participants missed the negation in the question), given its early placement, the perceived complexity of the question, and the longer instructional video in the structural condition (8.54 minutes vs. 3.58 minutes). Looking across all manipulation-check items, the overall validation results still indicate that the structural framing shaped participants’ conceptual understanding as intended. Complete item-level descriptive statistics and test results are provided in Appendix \ref{manipulation-appendix-details}.\looseness=-1

\subsection{RQ1: How Do Different Mental Models Affect the Way Users Exert Control in Their Writing Process?}

We first investigated how participants with functional or structural mental model priming exercised control when interacting with the system. Below, we report the main findings; where needed, additional details on the statistical analyses are provided in the Appendices. 

\paragraph{We observed no significant differences in control behavior-related measures across conditions.}

To evaluate control behavior, we analyzed 
interaction data that captured how
participants engaged with the \ac{AI}-based suggestions (described in Section \ref{sec:data-analysis} and outlined in Table \ref{tab:hypotheses-measures}), including the suggestion requests, the acceptance ratio, the edit ratio, the instruction count, and the user character ratio.

\begin{figure*}[!t]
\centering
\includegraphics[width=0.95\linewidth]{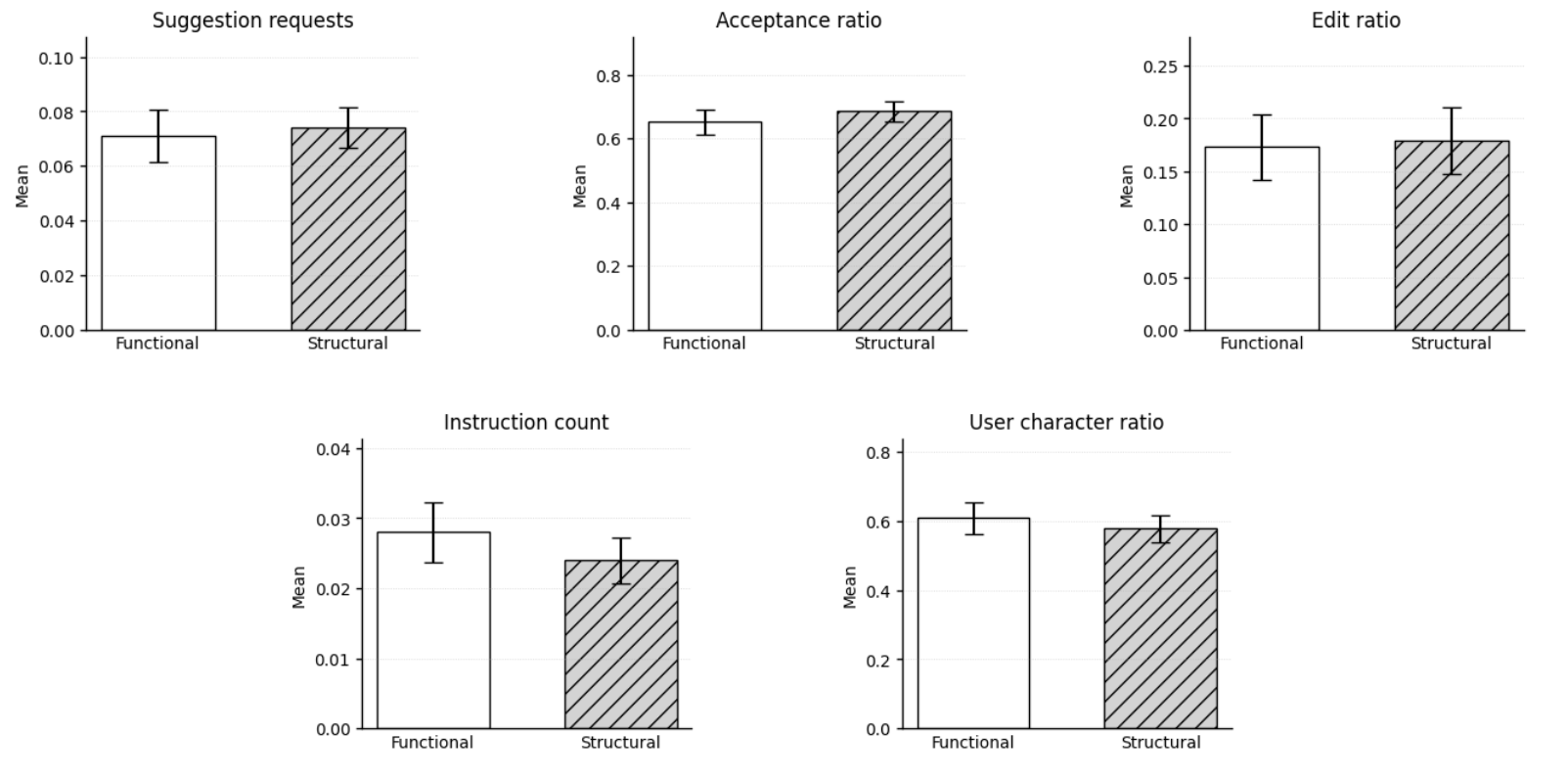}
\caption{Bar graphs of mean values (with error bars) for five control behavior measures---suggestion requests, acceptance ratio, edit ratio, instruction count, and user character ratio---across the functional and structural mental model conditions.}
\Description{Figure 3 shows five bar graphs with error bars comparing mean values of control behavior measures across functional and structural mental model conditions. The measures include calibrated suggestion requests, acceptance ratio, edit ratio, instruction count, and user character ratio. Each graph compares the functional condition (white bars) and the structural condition (striped gray bars). Across all five measures, the mean values between conditions are similar, and no statistically significant differences are observed.}
\label{fig:interaction-measures}
\vspace{-6pt}
\end{figure*}

Contrary to our hypotheses (H1a–H1d), structural participants did not request, accept, or edit suggestions more frequently, nor did they issue more instructions, and consistent with H1e, both groups contributed a similar amount of original text. As shown in Figure~\ref{fig:interaction-measures}, there were no statistically significant differences between the functional and structural conditions when looking at suggestion requests ($p = 0.409$), acceptance ratio ($p = 0.253$), edit ratio ($p = 0.869$), instruction count ($p = 0.726$), and user character ratio ($p = 0.305$). User character ratio and acceptance ratio met normality assumptions and were analyzed using independent samples $t$-tests; all other measures were evaluated using Mann--Whitney $U$ tests. Additional details about the descriptive and statistical analysis are in Appendix \ref{controlbehavious-appendix-details}.

Qualitative interview responses provide cues to why we did not observe differences across conditions for different behavioral measures. 
Across both conditions, participants described using \ac{AI}-generated suggestions opportunistically---most commonly when they \textit{``did not know what [they] were going to say''} (R5-F) or when they needed to \textit{``link different''} ideas together (R26-S). Some used suggestions primarily for inspiration and \textit{``guidance''} (R4-S) without selecting them, while others selectively adopted only specific elements such as a single \textit{``linking word''} (R17-F); in some cases, suggestions were accepted as-is. Thus, suggestion acceptance decisions seem guided by stylistic fit (e.g., \textit{``sounded like something I would write,''} R1-F) and local content relevance (e.g., \textit{``how relevant it was and how well it connected to the part of the sentence I had typed already,''} R32-S), rather than by participants’ understanding of how the system generated the suggestions.

Participants' views on the quality of the suggestions were similarly mixed across the two conditions.
Some found the suggestions helpful and appropriately toned (e.g., \textit{``helped me to form phrases that were appropriate to the context easily,''} R41-F), while others described them as \textit{``generic''} (R16-S), repetitive (e.g., \textit{``the suggestions were the same, with words placed differently,''} R3-F), or \textit{``basic and as expected''} (R10-S). All participants noticed that the suggestions contained grammatical and spelling errors. Experiences with the \textit{Instruction Box} were similarly mixed and did not differ by condition. Some participants appreciated that they \textit{``could write anything''} in the Instruction Box (R11-F), while others preferred having \textit{``a little more structure''} (R21-F). While some found the Instruction Box \textit{``useful''} in shaping suggestions (R32-S), others reported limited or inconsistent effects, sometimes needing to provide \textit{``a lot of instructions''} (R23-F) before seeing meaningful changes. This suggests that perceptions of suggestion quality and the Instruction Box’s effectiveness were primarily driven by participants’ personal writing-style and meta-prompting preferences, rather than by differences in mental models.

\paragraph{The types of instructions participants provided were similar across conditions, though participants in the structural condition wrote more descriptive instructions.}

To examine how participants used written instructions, we coded all instructions along three dimensions: \textit{tone/style shift}, \textit{qualifications articulation}, and \textit{structure prompting} (see Section \ref{ss:quant} for a description of our approach). Quantitative comparisons then showed no statistically significant differences in the types of instructions participants wrote across conditions. Specifically, chi-squared tests revealed no reliable association between condition and \textit{tone/style shift} ($\chi^2$(1, $N$ = 315) = 0.54, $p$ = 0.464), \textit{qualifications articulation} ($\chi^2$ = 1.30, $p$ = 0.254), or \textit{structure prompting} ($\chi^2$ = 0.06, $p$ = 0.802). Similarly, binomial tests, which we conducted to examine whether participants included some specific types of instructions more or less often than chance across the conditions, showed no deviation from a 50/50 distribution for \textit{tone/style shift} ($p$ = 0.851), \textit{qualifications articulation} ($p$ = 0.840), or \textit{structure prompting} ($p$ = 0.590). Full instruction-level coding distributions and statistical test results are reported in Appendix \ref{instruction-coding-appendix}. 

While the coding of the instruction categories did not reveal significant differences between conditions, closer qualitative examination of the instruction content shows some variation in how participants articulated tone and relevancy-related guidance. Overall, participants in the structural condition tended to provide more descriptive and context-rich instructions such as \textit{``Emphasize and weave my roles in higher education, mental health, and clinician training''} or \textit{``highlight using positive language that I am motivated and detail-oriented.''} In contrast, participants in the functional condition used brief keyword-style cues such as \textit{``use positive language''} or \textit{``highlight digital skills.''} The difference between the conditions was modest and on the the order of several cases, but reflected more elaborate instruction phrasing in the structural condition. Given that interview responses, across both conditions, revealed substantial variation in how participants experienced and used the Instruction Box (as explained earlier), it is not clear that these differences in descriptiveness of instructions are related to participants' understanding of the system. Nonetheless, the qualitative differences observed indicate the the relationship between mental models and instruction use needs to be further studied. 

Overall, both our quantitative and qualitative data suggest that participants’ moment-by-moment writing needs, perceived suggestion quality, and stylistic preferences primarily shaped how they exerted control during the writing process, with mental models playing a more limited role. 

\subsection{RQ2: How Do Different Mental Models Held by Users Influence the Quality of the Final Written Output?}

To investigate whether different mental model framings influenced the quality of participants’ final cover letter, we assessed writing quality via two methods (i.e., grammatical correctness and qualitative quality assessments, see Table \ref{tab:hypotheses-measures} for an overview). Based on our hypotheses, we expected that participants with a structural mental model would produce higher-quality cover letters, as judged via both automated grammar checks and qualitative quality assessments.  Full results for grammatical correctness, erroneous suggestion acceptance and cover letter quality analyses are reported in Appendices \ref{gramm-error-appendix} and \ref{cover-letter-quality-appendix}.\looseness=-1

\begin{figure*}[!t]
    \centering
    \includegraphics[width=1\textwidth]{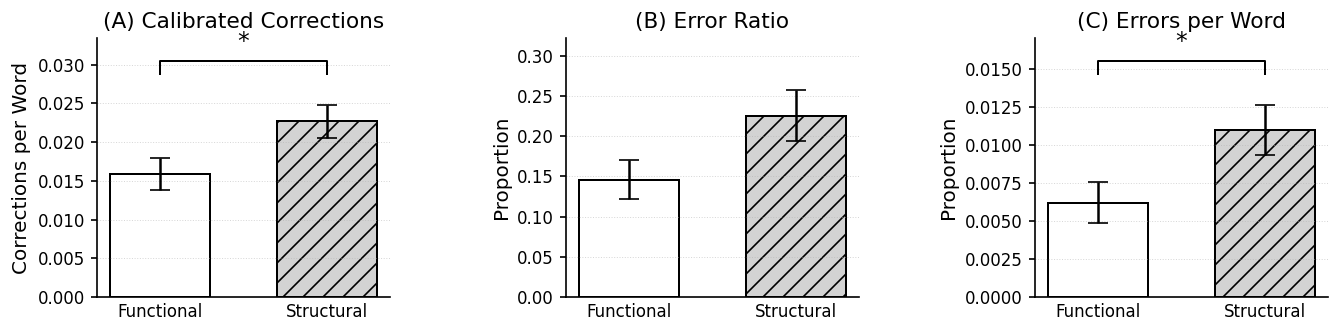}

    \caption{Bar graphs of mean values, with error bars, for calibrated corrections identified by Grammarly in final letters, error ratio, and accepted erroneous suggestions, comparing the functional and structural mental model conditions. An asterisk (*) denotes statistically significant differences ($p < 0.05$) for calibrated corrections and errors per word.}

    \Description{Three bar graphs with error bars comparing functional (white bars) and structural (striped gray bars) mental model conditions. (A) Calibrated corrections per word measure shows higher values in the structural condition, marked significant. (B) Error ratio measure also trends higher for structural but not significant. (C) Errors per word measure shows higher values in the structural condition, marked significant.}
    \label{fig:grammar-corr}
\end{figure*}

\paragraph{Participants in the structural condition wrote letters with more grammatical, punctuation, and spelling errors}
To evaluate grammatical correctness and surface-level writing quality (H2a), we used Grammarly's count of correctness suggestions, which captures the number of grammatical, punctuation, and spelling errors in each letter. \footnote{For brevity in this section, we use “grammatical errors” to refer to the combined count of grammatical, punctuation, and spelling errors captured in Grammarly's correctness score.} Contrary to our expectations, a one-sided independent samples t-test revealed a statistically significant difference in the \textit{calibrated corrections} identified using the Grammarly platform---i.e., the number of corrections suggested by Grammarly divided by word count---$t(46) = -2.316$, $p < 0.05$, with a moderate effect size ($d = 0.67$). This indicates that participants in the structural condition produced final letters with more grammatical errors compared to those in the functional condition. To better understand why the number of corrections required was higher for the structural condition, we conducted an exploratory analysis of how participants treated erroneous suggestions. As shown in Figure~\ref{fig:grammar-corr} and described next, participants in the structural condition accepted more suggestions containing grammatical errors and had a higher error ratio in accepted suggestions compared to those in the functional condition.

\begin{figure*}[t]
    \centering
    \includegraphics[width=0.50\textwidth]{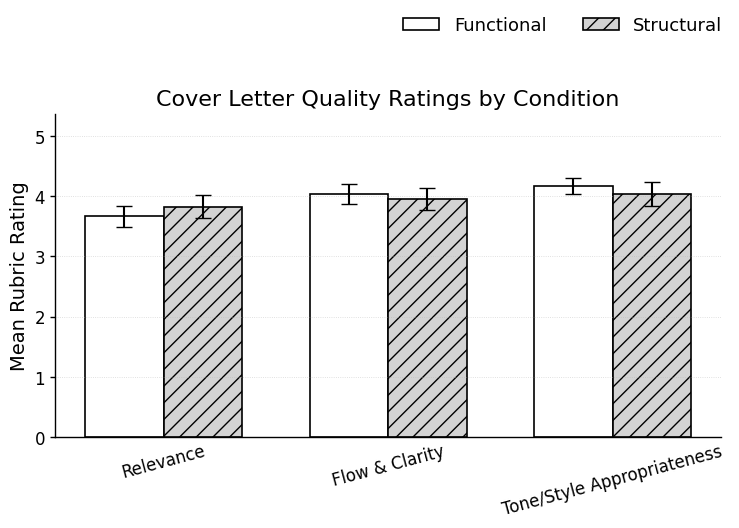}
    \Description{Bar graph showing mean rubric-based cover letter quality ratings, with error bars, across the dimensions of relevance, flow, and tone/style for two conditions. Ratings are high for both the functional (white bars) and structural (striped gray bars) mental model conditions, with similar values across all dimensions.}

    \caption{Bar graph of mean values (and error bars) for rubric-based cover letter quality ratings across dimensions of relevance, flow, and tone/style for the functional and structural conditions.}

    \label{fig:ext-eval-cover}
\end{figure*}

\paragraph{Participants in the structural condition accepted more erroneous AI suggestions} 

To better understand the discrepancy in cover letter quality, we conducted an exploratory analysis of the interaction data. This revealed a significant difference: participants in the structural condition accepted more suggestions containing grammatical errors (Figure~\ref{fig:interaction-measures}). A Mann–Whitney U test showed that the total number of accepted succestions \textit{errors per word} was significantly higher in the structural condition ($U = 170.0$, $p  < 0.05$, $r = .352$). The \textit{error ratio}, defined as the proportion of accepted suggestions containing grammatical errors, was also higher in the structural condition, though this difference was only marginally significant ($U = 204.5$, $p = .084$, $r = .250$). 
Because this analysis was exploratory 
we did not apply corrections for multiple comparisons ~\cite{Bender2001-pj}.\looseness=-1

Given the marginally imbalanced distribution of participants whose self-reported native language was English across mental model conditions, we explored whether being a native language speaker moderated the effect of mental model prompting. We conducted two-way \ac{ANOVA}s for both the \textit{error ratio} and \textit{accepted suggestions with errors (calibrated)}. Although these variables did not meet the Shapiro--Wilk normality assumption, Levene’s tests indicated homogeneity of variance, justifying \ac{ANOVA} use. No significant interaction effects were observed between mental model condition and native language for either the error ratio ($F(1, 44) = 0.408$, $p = .526$) or accepted suggestions with errors per word count ($F(1, 44) = 0.355$, $p = 0.554$).  No significant main effects of native language were observed for either outcome measures ($p = 0.216$ (error ratio) and $p = 0.258$ (accepted suggestions with errors)). \looseness=-1

From the semi-structured interviews and post-task surveys, we found that most participants across both conditions noticed grammatical errors in the \ac{AI}-generated suggestions, but adopted different strategies for handling them. Some participants reported editing errors after accepting suggestions (e.g., \textit{``I think once or twice, I did take one and then just remove the part that was wrong,''} R36-S), others avoided selecting erroneous suggestions entirely (e.g., \textit{``I felt less inclined to choose the ones with the errors,''} R19-F). Based on the observed \textit{error ratio} and \textit{errors per word} values, it appears that some participants may have accepted erroneous suggestions without subsequent changes, although no participant explicitly described this behavior. These strategies were observed in both mental model conditions, suggesting that participants relied on individual techniques to manage suggestion quality rather than following a condition-specific approach. As such, the qualitative data on handling grammatical errors does not conclusively explain why participants in the structural conditions have cover letters with higher-level grammatical errors. One possible interpretation is that the decision to accept a suggestion was often guided by writing style and goal, and that participants in the structural condition paid more attention to the suggestion content than to surface-level errors.  

\paragraph{Qualitative quality assessment of letter quality indicates no significant differences across groups}
Contrary to H2b, qualitative quality evaluation of the cover letters (described in Section \ref{sec:data-analysis}) revealed no statistically significant differences in the overall writing quality between the structural and functional conditions (Figure~\ref{fig:ext-eval-cover}). Mann–Whitney U tests yielded non-significant results for all three evaluation criteria: \textit{relevance} ($p = 0.398$), \textit{flow and clarity} ($p = 0.869$), and \textit{tone/style appropriateness} ($p = 0.954$). These findings indicate that, aside from grammatical errors, participants’ final letters were judged to be of comparable quality across the mental model conditions.

 Reflecting on RQ2, we observed that participants in the structural condition accepted more erroneous suggestions and produced letters with more grammatical errors, yet these differences were not reflected in the qualitative quality assessment of cover letter writing for factors such as relevance, fluency, and clarity, or tone/style appropriateness. This pattern indicates that a structural mental model may shape users’ attention or inattention to AI output---particularly in the presence of deliberate errors---without necessarily improving or degrading broader aspects of writing quality.
 
\subsection{RQ3: How Do These Mental Models Shape Users’ Overall Experience of Writing With the AI-Based Writing Assistant?}
For H3a–H3d, we hypothesized that participants in the structural condition would report higher ratings of perceived quality, control, ownership, and usefulness of the system. With the exception of ease of use, where participants with a structural mental model gave significantly higher ratings, we found no statistically significant differences between conditions. Across both groups, participants reported a strong sense of ownership and control over the final letter, along with generally high ratings of quality and usefulness. Because we tested five related subjective experience measures (perceived quality, control, ownership, usefulness, and ease of use), we applied a Bonferroni correction, yielding a corrected significance threshold of $\alpha = .01$. Figure~\ref{fig:quality_ratings} summarizes selected quantitative results, and detailed analyses are provided in Appendix \ref{post-task-appendix}.

\paragraph{Structural understanding led to more positive perceptions of CoAuthor’s usefulness and ease of use.}

We hypothesized that participants with a structural mental model would report a better experience (H3a), based on prior work suggesting that deeper system understanding supports user satisfaction and usability \cite[e.g.,][]{Kulesza2013-by}. Our findings provide only tentative support for this hypothesis.

Participants in the structural condition rated CoAuthor as easier to use than those in the functional condition. A one-sided independent samples $t$-test revealed a difference in \textit{ease of use}, $t(46) = -2.07$, $p < .05$, with a moderate effect size ($d = 0.60$). This effect was statistically significant at the uncorrected $\alpha = .05$ level, but did not meet the Bonferroni-corrected threshold of $\alpha = .01$. \textit{Perceived usefulness} was assessed using a four-item composite scale with acceptable internal consistency (Cronbach’s $\alpha = 0.76$). Although the perceived usefulness score was on average slightly higher for the structural condition, the difference between conditions on this composite score was not statistically significant ($t(46) = -0.78$, $p = .220$. Additionally, none of the individual item comparisons reached statistical significance. These comparisons are visualized in Figure~\ref{fig:quality_ratings}.

Qualitative interview data provided additional context for when participants' perspectives appear to overlap and when they diverge. 
Across both conditions, participants described CoAuthor as useful for improving phrasing, overcoming writer’s block, generating ideas, and accelerating the writing process, with many completing a full draft within 20 minutes. Several also found the tool helpful for structuring their letters. At the same time, participants in both conditions raised similar concerns about generic or robotic-sounding suggestions and occasional grammatical errors. 
These shared perceptions of their interactions with the platform might explain why perceived usefulness ratings remained similarly high across both conditions.

Where the conditions diverged was in how participants' self-reports linked their understanding of the system with their confidence and perceived sense of control when using the platform.
A few participants in the functional condition expressed that limited insight into how the system worked, constrained their trust and sense of control (e.g., \textit{``I need to know where it is getting the information from,''} R21-F). In contrast, some participants in the structural condition explicitly linked their confidence to the explanatory onboarding (e.g., \textit{``[The platform] works well and it’s very functional \ldots I am able to either use the suggestions effectively or dismiss them if they are bad,''} R2-S). 
Together, these participants' comments provide further insights into why the structural understanding might have only selectively improved perceptions of ease of use, while the more quantitative evaluations of usefulness were minimally higher for the structural condition.

\paragraph{We found no significant differences in perceived letter quality, control, or ownership across conditions.}
Contrary to H3b--H3d, quantitative analyses revealed no statistically significant differences between the functional and structural conditions in participants’ perceived letter quality, control, or ownership. Participants in the structural condition rated CoAuthor’s suggestions as slightly more error-free than those in the functional condition, although this difference was not statistically significant ($t(46) = -1.48$, $p = .072$, $d = 0.43$). They also reported marginally higher ratings for the final letter’s grammatical correctness, tone, and content, but these differences were likewise non-significant ($p = .801$ and $p = .134$, respectively). Perceived ownership was analyzed using a composite score across items (Cronbach’s $\alpha = .714$), whereas perceived control was assessed using multiple items but is reported at the item level due to lower internal consistency ($\alpha = .658$). Both conditions reported similarly high levels of perceived control and ownership. Overall, we found no statistically significant differences across quality-, control-, or ownership-related measures. Mean ratings for all self-reported items are shown in Figure~\ref{fig:quality_ratings}. Additional details are included in Appendix \ref{post-task-appendix}.

\begin{figure*}[!t]
    \centering
    \includegraphics[width=1\textwidth]{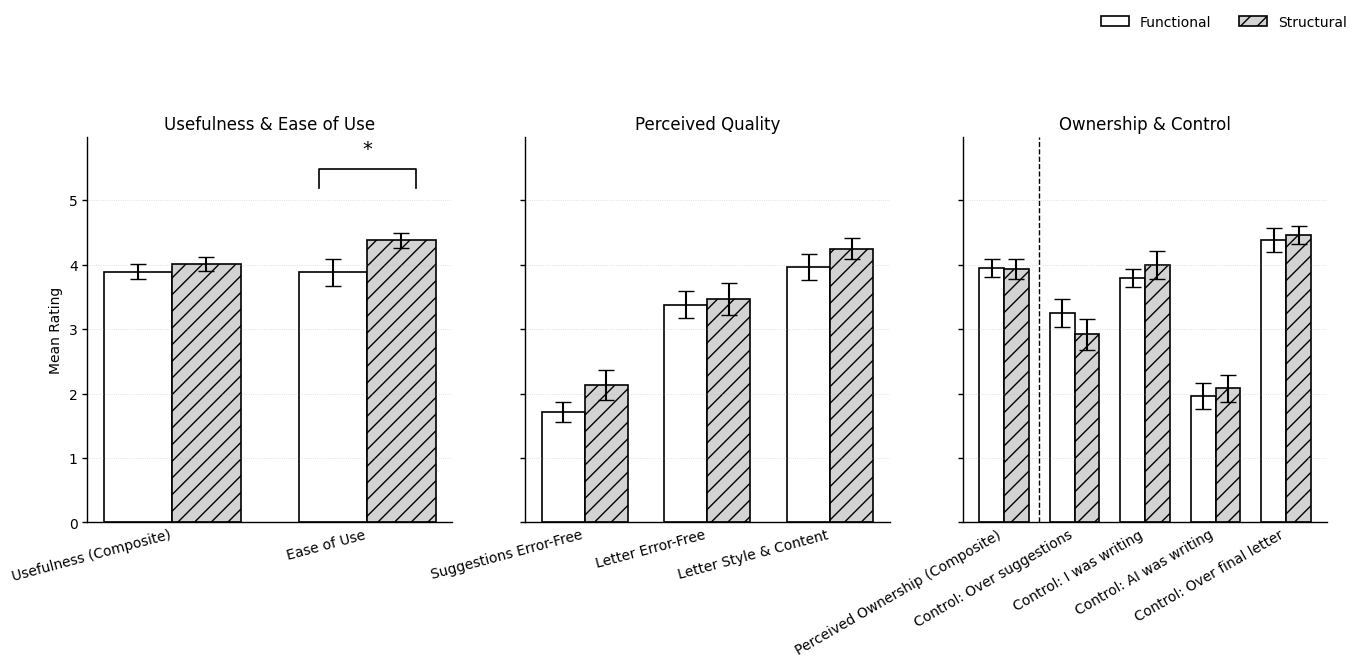}
    \caption{Bar graphs of mean values (and error bars) for self-reported ratings across conditions for usefulness, ease of use, perceived quality, perceived ownership, and perceived control. A one-sided independent samples $t$-test revealed a difference in \textit{ease of use}, with a moderate effect size ($d = 0.60$). As marked on the left-most plot, this effect was statistically significant at the uncorrected $\alpha = .05$ level, but did not meet the Bonferroni-corrected threshold of $\alpha = .01$.}
    \Description{Three grouped bar graphs with error bars comparing functional (white bars) and structural (striped gray bars) conditions on self-reported ratings. On the left, the first graph shows usefulness (composite) and ease of use, with structural condition rated significantly higher for ease of use (*) at the uncorrected alpha level ($\alpha = .05$). The second graph shows perceived quality, including suggestions being error-free and letter style and content, with structural condition values being slightly higher but with no significant differences. The third graph shows ownership and control measures, including perceived ownership (composite) and four control dimensions, with similar ratings across conditions.}
    \label{fig:quality_ratings}
\end{figure*}

\paragraph{Interaction features, not mental models, appear to anchor participants' perceptions of quality, control, and ownership.}

Qualitative findings help explain why perceived quality,  control, and ownership remained consistently high across both mental model conditions. Overall, participants noted that the modified CoAuthor platform felt different from other forms of \ac{AI}-based writing assistance. Many participants remarked that they felt a high level of control when using CoAuthor, describing it more as a supportive ``friend'' offering help (e.g., \textit{``I was basically just writing something and asking a friend how would you help me describe this,''} R3-F) rather than a tool that did all the work for them (e.g., \textit{``I like [that] you still have to do the work [...]. For [tools like] ChatGPT [...], it just writes the whole thing and you don't really do anything. This [...] you still having to do it yourself,''} R8-S). 

Participants identified several concrete reasons for experiencing a high sense of control. The most common factors were that they could request suggestions whenever they wanted (e.g., \textit{``I could choose when to ask for suggestions really easily,''} R24-S), directly edit the text in both the editor and the \textit{Instruction Box} (e.g., \textit{``I always had the ability to go back and erase what I didn't like,''} R17-F), and choose whether or not to accept the suggestions (e.g., \textit{``I could reject all of them or I could choose an option,''} R27-F). Moreover, participants noted that they felt in control because CoAuthor did not nudge or pressure them to use its features (e.g., \textit{``I like that I have to ask it for the suggestion,''} R5-F)---this observation was coded four times more often in the functional condition compared to the structural condition. Participants further emphasized that they felt a sense of control because the suggestions were relevant to what they had written in the \textit{Instruction Box} and text editor (e.g., \textit{``it gave relevant suggestions that helped me express my inner words and I just put my feelings in the instruction box,''} R34-S). Additionally, the short length of the suggestions---typically less than a full paragraph---contributed to the sense of perceived control, as it required them to do more of the writing themselves. In the same vein, the fact that users had to enter one instruction at a time encouraged a more active and hands-on writing process. Notably, several participants across both conditions highlighted points where they felt some loss in control. Few participants in functional condition expressed that they lost some control when they \textit{``just kept asking for suggestions''} (R1-F), often because they ran out of ideas or were under \textit{``time constraint''} (R33-F). On the other hand, a participant in the structural condition expressed that having the ability to put in \textit{``more instructions [at single use of the Instruction Box]''} (R18-S) would increase their sense of control.

Participants primarily defined ownership in three ways. First, they described ownership as having their own ideas reflected in the writing (e.g., \textit{``writing reflects my thoughts, my arguments''} R9-F). Second, some associated ownership with doing the writing themselves (e.g., \textit{``I have really done the job of writing it myself,''} R20-S). Third, others emphasized authenticity and personal voice (e.g., \textit{``it would have to be genuine and honest,''} R14-S). Along these lines, participants suggested that a sense of ownership often came from feeling proud of what they had written. These conceptualizations of ownership were present across both mental model conditions. Several participants also noted that one could still maintain a sense of ownership when using help from others or \ac{AI}-based writing assistance, depending on how much help was used and whether it aligned with these core criteria.
Most participants expressed a strong sense of ownership because they felt in control---they could write their own text, were not nudged, could choose whether or not to accept suggestions, and could request suggestions whenever they wanted. Many also felt ownership because they were able to bring in their own ideas and personal experiences. Several participants reported a strong but incomplete sense of ownership over the writing, with seven characterizing their contribution at approximately 75\%--85\% (3 in the structural condition, and 4 in the functional one; e.g., \emph{``If I were to give it a percentage, I’d probably say around 80--85\%, because it's my experiences, but in terms of how the words are formulated, CoAuthor helped me formulate most of the words,''} R35-F).
However, some participants appeared conflicted. They reported a weaker sense of ownership when the writing did not feel authentic, did not reflect their personal style, or when they did not feel proud of the final output. When participants accepted too many suggestions, they found the draft letter felt less personalized. These observations were consistent across the conditions. 

Taken together, these qualitative accounts help contextualize and explain the quantitative patterns observed in participants’ overall experience of writing with the system. Regardless of participants’ mental model of how the system worked, perceptions of quality, usefulness, control, and ownership remained largely comparable across conditions, reflecting the central role of interaction features and affordances that supported user choice, idea contribution, and iterative editing. The qualitative data suggest that these affordances enabled participants to feel similarly empowered and satisfied with the system’s outputs, even when their conceptual understanding of the system differed. In contrast, perceived ease of use varied more directly with participants’ understanding of how the system operated, as clearer mental models reduced uncertainty about available actions and system behavior, making interaction feel more intuitive and less effortful.
\looseness=-5

\section{Limitations and Positionality}

This study has several limitations that shape how its findings should be interpreted. First, while we aimed to elicit distinct mental models of the \ac{AI}-based writing assistant through carefully crafted videos providing functional and structural descriptions of the system, these framings represent static depictions of a user’s mental model at a single point in time. In practice, mental models are dynamic and may evolve with each interaction. Despite strict inclusion criteria targeting participants with limited prior experience, many participants likely entered the study with pre-existing beliefs about \ac{AI} systems that could have influenced their perceptions and behaviors. Furthermore, creating a truly structural mental model of a system as complex and opaque as an \ac{AI}-based writing assistant is inherently challenging. We took deliberate steps to scaffold this understanding, but the extent to which participants internalized the structural framing likely varied. Another limitation concerns participant variability, particularly in writing proficiency. Although we attempted to recruit participants with comparable experience levels and provided a consistent task, underlying differences in writing skill may have shaped engagement with the system and perceptions of suggestion quality. 
In addition, while the prevalence of participants whose native language was not English did not differ significantly across conditions, it was marginally imbalanced and may have introduced subtle variability in how participants interpreted and evaluated \ac{AI}-generated writing.

Our research team’s positionality also influenced the framing and interpretation of this work. The five researchers involved span both academic and industry settings in the Global North (Canada and the U.S.), with expertise in natural language processing, social computing, human–computer interaction, human–robot interaction, responsible AI, and system safety. These disciplinary perspectives shaped the study’s focus on user control, ownership, and safe system interaction. We acknowledge that our interpretations are informed by this interdisciplinary but Western-centric vantage point, and that broader perspectives---particularly those situated outside of North America or grounded in alternative epistemologies---are critical for expanding this line of inquiry. \looseness=-1

\section{Discussion}
In this study, we explored how users’ mental models of an \ac{AI}-based writing assistant shaped interaction behavior, final writing quality, and overall experience. Our manipulation was effective in that participants in the structural condition demonstrated a deeper conceptual understanding of how the system operated and performed better on factual knowledge questions. At the same time, many downstream outcome measures showed few or no statistically significant differences between conditions. Participants across both conditions displayed similar control behavior and produced letters that were similar in tone, clarity, and relevance. In addition, both groups reported a comparably strong sense of quality, ownership and control. These findings suggest that differences in users’ conceptual understanding of the system do not straightforwardly translate into broad differences in writing quality or perceived experience. 
Where differences did emerge, they should be interpreted cautiously. Participants in the structural condition showed a consistent trend toward higher perceived ease of use and greater acceptance of erroneous suggestions, while also producing letters with a higher number of grammatical errors. Rather than indicating strong causal effects of structural understanding, these patterns suggest a subtle shift in how participants related to and used the system. Drawing on prior work on human–machine trust and reliance, one plausible interpretation is that structural framing functioned as a cue of system competence, fostering trust and shaping reliance tendencies, including greater acceptance of deliberately inserted erroneous suggestions. \cite{Romeo2025_automationBiasReview,Zhai2024-fa,Ng2025-rs,Vered2023-ud}. In this sense, overtrust is not presented here as a definitive explanation, but as a useful lens for interpreting why modest increases in ease of use may co-occur with a greater tendency to accept flawed suggestions and letters with more grammatical errors.

\paragraph{Re-assessing the link between mental models and safe control in some human–AI interaction contexts.}
In traditional safety-critical systems, “hazardous situations” are often well-defined events with known consequences—such as a plane deviating from its flight path or a robot misinterpreting a command \cite{Leveson2018-no}. In \ac{AI}-based writing assistants, hazards are subtler and more subjective. They may involve insertion of specific ideas via AI-generated text, misrepresentation of cultural content, or uniformization of writing styles \cite{Li2024-jm, Poddar2023-ki, Jakesch2023-tx}. Additionally, these risks are best understood as emerging from patterns of interaction over time rather than as isolated, single-point failures \cite{Rismani2025-ae,Kirk2025-bp, Wallach2025-db}. This implies that monitoring for such hazards, requires oversight mechanisms over the course of these interactions. 

One common assumption in system safety is that a more accurate mental model should support more effective oversight and intervention \cite{Gaspar2020-xj,Leveson2012-jp}. Our findings complicate this narrative within the context of AI-assisted writing systems with preconfigured erroneous AI outputs. Participants with a structural mental model did not consistently intervene to correct erroneous AI-generated suggestions and, if anything, tended to accept such suggestions more readily, resulting in letters with a higher number of grammatical errors. At the same time, they also reported the system as easier to use. This points to a key tradeoff that might be especially salient in lower-stakes, productivity-oriented domains like everyday writing where users may prioritize efficiency, fluency, and reduced cognitive effort over careful verification of each system output. From this perspective, what appears as reduced oversight may in fact just reflect a reasonable adaptation to task demands---e.g., optimizing efficiency over producing a polished writing outcome---rather than a failure of understanding.

In settings where users might hold distinct conceptualizations of what constitutes a system error, failure, or harm, system designers may need to shift focus. Rather than providing users with a more in-depth description of how a system works, emphasis should be placed on scaffolding good interaction affordances: mechanisms that surface suggestion quality, prompt reflection, or highlight alternative options \cite{Kadoma2024-sv, Buschek2021-xh}. This resonates with system safety’s emphasis on designing feedback-rich environments that allow users to recognize system states, monitor their own actions, and recover from errors \cite{Dobbe2022-kf}. Such design strategies may better support writers in exercising meaningful and context-sensitive control, particularly in situations where what counts as a “safe” action is not obvious.

\paragraph{Rethinking mental model attunement in designing \ac{AI}-based writing systems.}
While our findings suggest that a more in-depth structural mental model does not necessarily lead to better control or writing quality, this does not diminish the value of understanding how mental models shape AI use. In fact, our data show that mental models still meaningfully influence how users approach interaction, perceive system usability, and articulate their strategies for directing AI behavior. For example, participants in the structural condition rated the system as significantly easier to use and were more descriptive in customizing tone or intent through the Instruction Box. These behaviors suggest that structural mental models can scaffold more purposeful interaction, even if they do not guarantee improved outcome quality.\looseness=-1 

Importantly, \ac{AI}-based writing systems are not static and non-deterministic. They evolve rapidly through updates, retraining, and interaction data, making it difficult to determine what a “correct” or “complete” mental model even looks like. From a systems perspective, this introduces new challenges for maintaining safe user-system interaction: mental models may quickly become outdated, incomplete, or misaligned with current system behavior \cite{Mehmood2025-jh, Wang2025MentalModelsGenAI}. From this perspective, mental models should not be treated as fixed educational outcomes, but as evolving resources that are continuously renegotiated through use. This has important design implications. Rather than relying on one-time instructional framing, systems should support ongoing recalibration through adaptive scaffolding \cite{Passi2025-qp}. This may include progressive disclosure of system behavior \cite{Cortinas-Lorenzo2025-dx}, just-in-time explanations tied to specific outputs \cite{Swaroop2025-ie}, lightweight uncertainty indicators \cite{Schafer2025-pr}, and interaction techniques that encourage users to periodically reflect on whether AI suggestions align with their goals \cite{Ma2025-dd}. Supporting better mental models, in this sense, is less about transmitting backend technical detail and more about enabling sustained, critically informed engagement over time.

\paragraph{Importance of affordances versus user understanding in mitigating potential sources of sociotechnical harms.}
In safety-critical systems, ensuring human control is paramount; users must be able to monitor, override, and intervene in ways that prevent harm. However, in creative or productivity-oriented domains like writing, the need for user control is more context-dependent. It depends on the nature of the writing task, the goals of the writer, and broader contextual factors such as whether \ac{AI} assistance is permitted or desirable \cite{Gero2023-du, Draxler2024-ut}. Our findings suggest that participants experienced a strong sense of control and ownership when using CoAuthor, regardless of mental model framing. Notably, CoAuthor offered more flexible interaction affordances than many typical writing assistants: users could freely request suggestions, revise them, or reject them entirely. This flexibility appeared to support a high baseline sense of control and ownership  \cite{Kadoma2024-sv}. Compared to the many \ac{AI} tools that generate long outputs or push users toward passive acceptance, the modified CoAuthor’s interaction design may be a key reason participants reported feeling like authors, not just editors. This raises an important design implication: if the goal is to support user control in writing assistants, the priority may not necessarily be better system understanding, but better affordances.\looseness=-1  

\paragraph{Scope, generalizability, and baseline framing.}
Our findings are grounded in scenarios where the system deliberately produced observable errors in its suggestions. This design choice allowed us to directly examine whether users could detect and intervene in flawed outputs, but it also limits the generalizability of our conclusions. It remains an open question whether similar trust, reliance, and control dynamics would emerge under conditions in which no errors were deliberately introduced or in other forms of writing support such as ideation, summarization, or stylistic rewriting. Structural mental models may play different roles in such settings, for example by shaping expectation management rather than error detection.

At the same time, our investigation of user control was not limited to error correction alone. Through analysis of how participants used the Instruction Box to adjust tone, structure, and intent, we also observed broader forms of intervention through which users attempted to steer and personalize the writing process. These behaviors suggest that even when users show tendencies toward a lack of effective oversight at the level of individual suggestions, they may still engage in higher-level, directional forms of control over the overall output. This distinction is important for understanding ownership and control in AI-assisted writing beyond narrow notions of error detection.

\section{Conclusion}

In this study, we investigated how different mental model framings of an \ac{AI}-based writing assistant influence users’ behaviors, writing outcomes, and experiences. Participants exposed to structural explanations developed a deeper understanding of the system and reported higher perceived ease of use. However, they also showed a greater tendency to accept erroneous suggestions and produced cover letters with more grammatical errors. We posit that this pattern may reflect overtrust, whereby structural framing functioned as a cue of system competence, increasing reliance even when active oversight was warranted due to staged errors. Notably, participants across both conditions expressed a strong sense of ownership and control, largely fostered by the AI-based writing assistant's interaction features, which allowed users to actively write and edit. These results suggest that providing users with accessible means of control may be more impactful than deeper comprehension of the system alone when it comes to fostering a sense of ownership and control. Looking ahead, research should further examine how to cultivate appropriate mental models and interaction paradigms that empower users to exercise oversight when necessary.\looseness=-1

\begin{acks}
We gratefully acknowledge all participants for their time and contributions to this research. We also thank Gauri Sharma, Kiara Wimbush, and Timothy Ko Lee for their invaluable assistance in running the pilot studies and setting up the research platform.
\end{acks}
\bibliographystyle{ACM-Reference-Format}
\bibliography{main-base}

\clearpage
\appendix
\section{Demographics Survey}
\label{app:demographics}

This section lists the survey questions we administered to prospective participants in order to select and stratify participants for our study. 

\begin{enumerate}
    \item What is your age?\\
    \textit{Single line text}

    \item What is your gender?\\
    \textit{Single choice}
    \begin{itemize}
        \item Female
        \item Male
        \item Non-binary
        \item Prefer not to answer
        \item Other
    \end{itemize}

    \item What is the highest level of education you have completed?\\
    \textit{Single choice}
    \begin{itemize}
        \item Primary school
        \item High school
        \item Post-secondary vocational institution (trade and technical school) or CEGEP
        \item 3 or 4-year undergraduate university program
        \item Master’s degree program
        \item PhD program
        \item None of the above
        \item Prefer not to answer
    \end{itemize}

    \item How often do you write in English?\\
    \textit{5-point Likert scale}

    \item How would you rate your confidence in writing in English?\\
    \textit{5-point Likert scale}

    \item What is your native language?\\
    \textit{Single choice}
    \begin{itemize}
        \item English
        \item French
        \item Other
    \end{itemize}

    \item How would you rate your cover letter writing abilities?\\
    \textit{5-point Likert scale}

    \item How would you rate your level of confidence in writing cover letters?\\
    \textit{5-point Likert scale}

    \item Which one of the following AI-based writing assistants have you used?\\
    \textit{Multiple choice(can choose multiple)}
    \begin{itemize}
        \item ChatGPT
        \item Grammarly
        \item Wordtune
        \item Notion
        \item None of the above
        \item Other
    \end{itemize}

    \item How often do you use any of the AI-based writing assistants stated in the previous
question?\\
    \textit{5-point Likert scale}
\end{enumerate}

\section{System Understanding Survey}
\label{app:understanding}

This section lists the survey questions we used to probe participants understanding of the AI-based writing assistant platform after they watched the instructions videos for one of the experimental conditions they were assigned to. 


\begin{enumerate}
    \item CoAuthor is a writing assistant that does not rely on Artificial Intelligence (AI).
    \begin{itemize}
        \item True
        \item False
        \item I do not know.
    \end{itemize}

    \item CoAuthor offers phrases and sentences when a user requests suggestions.
    \begin{itemize}
        \item True
        \item False
        \item I do not know.
    \end{itemize}

    \item Where should the user write the main body of their cover letter on the CoAuthor platform?
    \begin{itemize}
        \item Text editor
        \item Instruction box
        \item Both
        \item I do not know.
    \end{itemize}

    \item CoAuthor will automatically generate suggestions when the user stops typing.
    \begin{itemize}
        \item True
        \item False
        \item I do not know.
    \end{itemize}

    \item CoAuthor filters and trims the original list of suggestions generated by the large language model before showing it to the user.
    \begin{itemize}
        \item True
        \item False
        \item I do not know.
    \end{itemize}

    \item CoAuthor gives suggestions based on what is a likely continuation of the last sentence in the text editor.
    \begin{itemize}
        \item True
        \item False
        \item I do not know.
    \end{itemize}

    \item The content of the Instruction Box is sent to a different language model than the text editor.
    \begin{itemize}
        \item True
        \item False
        \item I do not know.
    \end{itemize}
\end{enumerate}

\section{Post-task Survey}
This section lists the questions asked in the post task survey in order to assess participants' perceptions of their experience with the AI-based assistant. The responses were given on 5-point Likert scales. 
\label{app:posttask}


\begin{enumerate}
    \item Rate the following statements about the cover letter that was written using CoAuthor (5-point Likert Scale). 
    \begin{itemize}
        \item I am the main contributor to the content of the cover letter.
        \item I am accountable for all aspects of the cover letter.
        \item I feel like I am the author of the cover letter.
        \item I feel like the cover letter represents me.
    \end{itemize}

    \item Rate the following statements about the cover letter that was written using CoAuthor (5-point Likert Scale). 
    \begin{itemize}
        \item I am the main contributor to the content of the cover letter.
        \item I am accountable for all aspects of the cover letter. 
        \item I feel like I am the author of the cover letter. 
        \item I feel like the cover letter represents me. 
        \item I was able to customize the provided 
    \end{itemize}
    \item Rate the following statements about your experience of controlling CoAuthor when writing the cover letter (5-point Likert Scale). 
    \begin{itemize}
        \item The suggestions influenced the content of the cover letter. 
        \item I was able to customize the provided suggestions using the Instruction Box.  
        \item I felt in control of the quality and content of the generated suggestions.  
        \item I felt like I was writing the text and CoAuthor was assisting me. 
        \item I felt like CoAuthor was writing the text and I was assisting. 
        \item I felt in control of what was included in the final letter. 
        \item CoAuthor was easy to use.
    \end{itemize}
    \item Rate the following statements about the quality of the suggestions, the final letter, and CoAuthor (5-point Likert Scale).
    \begin{itemize}
        \item The suggestions were relevant.
        \item The suggestions were error-free.
        \item The suggestions were helpful.
        \item Customizing suggestions using the Instruction Box was helpful.
        \item The final letter is error-free.
        \item The final letter has the appropriate style and content for a cover letter.
    \end{itemize}
\end{enumerate}

\section{Cover Letter Assessment and Instruction Coding Rubrics}
\label{app:rubrics}

This section reports the assessment criteria we used to rate the overall quality of the cover letters (\S\ref{app:cover-letter}), as well as the coding rubric we used to assess the type of instructions participants provided in the instruction box (\S\ref{app:instruction-types}).

\subsection{Qualitative Quality Assessment Rubric}
\label{app:cover-letter}

Table \ref{tab:coverletter-rubric} describes the three criteria the research team used to score the cover letters. 
\begin{table}[h!t]
\centering
\scriptsize
\caption{Cover letter assessment rubric used by evaluators.}
\label{tab:coverletter-rubric}
\begin{tabular}{p{0.30\columnwidth} p{0.65\columnwidth}}
\toprule
\textbf{Criterion} & \textbf{Rating anchor (5 = Excellent), (1 = Poor)} \\
\midrule

Demonstration of relevant skills and experience &
The letter clearly highlights the candidate’s skills, knowledge, and experiences that directly relate to the job posting. Examples are used to support claims.
\\

Flow and clarity &
The flow of the letter is appropriate for a cover letter and the writing is clear. The content transitions smoothly from one idea to the next. Sentences and paragraphs are logically organized, making the letter easy to read and understand without confusion or abrupt shifts.
\\

Tone and style &
The writing has an appropriate and compelling tone and style for a cover letter. The letter has a professional tone and the candidate’s voice comes through as confident, enthusiastic, and sincere.
\\

\bottomrule
\end{tabular}
\end{table}

\subsection{Instruction Coding Rubric}
\label{app:instruction-types}

Table \ref{tab:instruction-rubric}  includes a brief description of the criterion that the research team used to annotate each one of the instructions that participants wrote. 

\begin{table}[h!t]
\centering
\scriptsize
\caption{Instruction coding rubric used to annotate user-written instructions.}
\label{tab:instruction-rubric}
\begin{tabular}{p{0.70\columnwidth} p{0.20\columnwidth}}
\toprule
\textbf{Criterion} & \textbf{Response} \\
\midrule

\textbf{Tone or style shift}\\
Tone is focused on the emotional stance and attitude the writer brings through in their writing. Style is broader than tone, and captures how the writer expresses their message using different words, unique voice, and structure. For example, ``use positive language'' shifts tone; ``make words professional'' or ``decrease monotony'' shifts style.
&
Yes / No
\\

\addlinespace
\textbf{Qualifications articulation}\\
Does the instruction articulate the candidate’s relevant qualities, qualifications, or skills? For example, ``add graduate degree at McGill University'' is specific, whereas ``highlight great asset'' is too broad.
&
Yes / No
\\

\addlinespace
\textbf{Structure prompting}\\
Does the instruction cue the structure of the cover letter (e.g., ending, section transition)? For example, ``conclusive sentence'' or ``write something about company.''
&
Yes / No
\\

\bottomrule
\end{tabular}
\end{table}

\section{Manipulation Check Details}
\label{manipulation-appendix-details}
This section includes further details about the manipulation check results. Table~\ref{tab:manipulation-appendix} provides full item-level descriptive statistics and nonparametric test results for all manipulation-check questions referenced in the main paper. It reports item-level descriptive statistics (mean, standard deviation, and median) for each manipulation-check question by condition, along with two-tailed Mann--Whitney U tests (U, Z, p). Items were coded as 1 = correct and 0 = incorrect; means therefore represent proportions correct. The final row (“Total”) corresponds to the summed manipulation-check score reported in the main paper.

\begin{table}[ht]
\centering
\scriptsize
\def\arraystretch{1.1}
\setlength{\tabcolsep}{0.4em}
\caption{Descriptive statistics and Mann--Whitney U tests for manipulation check questions by condition.}
\label{tab:manipulation-appendix}
\begin{tabular}{@{}c|ccc|ccc|cc|ccc@{}}
\textbf{Q} &
\multicolumn{3}{c|}{\textbf{Functional (MM = 1)}} &
\multicolumn{3}{c|}{\textbf{Structural (MM = 2)}} &
\multicolumn{2}{c|}{\textbf{Mean Rank}} &
\textbf{U} & \textbf{Z} & \textbf{p} \\
 & Mean & SD & Median & Mean & SD & Median & MM=1 & MM=2 &  &  &  \\
\hline
1 & 1.00 & 0.00 & 1.00 & 0.79 & 0.41 & 1.00 & 27.00 & 22.00 & 228.0 & -2.34 & .019 \\
2 & 1.00 & 0.00 & 1.00 & 0.96 & 0.20 & 1.00 & 25.00 & 24.00 & 276.0 & -1.00 & .317 \\
3 & 0.96 & 0.20 & 1.00 & 0.96 & 0.20 & 1.00 & 24.50 & 24.50 & 288.0 & 0.00 & 1.000 \\
4 & 1.00 & 0.00 & 1.00 & 1.00 & 0.00 & 1.00 & 24.50 & 24.50 & 288.0 & 0.00 & 1.000 \\
5 & 0.21 & 0.41 & 0.00 & 0.79 & 0.41 & 1.00 & 17.50 & 31.50 & 120.0 & -4.00 & <.001 \\
6 & 0.75 & 0.44 & 1.00 & 0.92 & 0.28 & 1.00 & 22.50 & 26.50 & 240.0 & -1.53 & .125 \\
7 & 0.25 & 0.44 & 0.00 & 0.67 & 0.48 & 1.00 & 19.50 & 29.50 & 168.0 & -2.87 & .004 \\
\hline
Total & 5.17 & 0.92 & 5.00 & 6.08 & 1.06 & 6.00 & 18.44 & 30.56 & 142.5 & -3.15 & .002 \\
\end{tabular}

\end{table}

\section{Control Behavior Measures Statistical Analysis Details}
\label{controlbehavious-appendix-details}
This section elaborates on the results from the statistical analysis completed for various measures for control behavior. Table \ref{tab:interaction-measures-full} reports the full descriptive and inferential statistics for the interaction measures summarized in Figure~\ref{fig:interaction-measures}. For each measure, we report group-level descriptive statistics for the functional and structural conditions, along with the statistical test used, corresponding test statistics, exact $p$-values, and effect sizes. Measures that met normality assumptions were analyzed using independent-samples $t$-tests, whereas measures that violated normality assumptions were analyzed using Mann--Whitney $U$ tests.

\begin{table*}[h!t]
\centering
\scriptsize
\def\arraystretch{1.1}
\setlength{\tabcolsep}{0.4em}
\caption{Full statistical results for interaction measures.}
\label{tab:interaction-measures-full}
\begin{tabular}{@{}lcccccc@{}}
\toprule
Measure &
Functional condition &
Structural condition &
Test &
Statistic &
$p$ &
Effect size \\
\midrule
Suggestion requests (calibrated) &
$M = 0.071$, $SD = 0.047$ &
$M = 0.074$, $SD = 0.036$ &
Mann--Whitney $U$ &
$U = 248$, $Z = -0.83$ &
.409 &
$r=0.12$ \\

Acceptance ratio &
$M = 0.651$, $SD = 0.188$ &
$M = 0.685$, $SD = 0.161$ &
$t(46)$ &
$t = -0.67$ &
.253 &
$d = -0.19$ \\

Edit ratio &
$M = 0.173$, $SD = 0.154$ &
$M = 0.179$, $SD = 0.153$ &
Mann--Whitney $U$ &
$U = 280$, $Z = -0.17$ &
.869 &
$r=0.02$ \\

Instruction count &
$M = 0.028$, $SD = 0.021$ &
$M = 0.024$, $SD = 0.016$ &
Mann--Whitney $U$ &
$U = 271$, $Z = -0.35$ &
.726 &
$r=0.05$ \\

User character ratio &
$M = 0.609$, $SD = 0.227$ &
$M = 0.578$, $SD = 0.187$ &
$t(46)$ &
$t = 0.51$ &
.305 &
$d = 0.15$ \\
\bottomrule
\end{tabular}
\end{table*}

\section{Instruction Coding Statistical Analysis Details}
\label{instruction-coding-appendix}
In this section, we provide more indepth information about two types of statistical analysis that we conducted to compare the instructions that the participants provided across the mental model conditions. Table \ref{tab:instruction-dimensions-chisq} overviews the distribution of codes for participants instructions across the functional and structural conditions. For each instruction dimension, the table shows the number of instructions in which the feature was absent (0) or present (1) for each condition, along with the results of chi-squared tests assessing whether instruction type was associated with experimental condition.

\begin{table}[ht]
\centering
\scriptsize
\def\arraystretch{1.1}
\setlength{\tabcolsep}{0.25em}
\caption{Distribution of coded instruction dimensions by condition (instruction-level).}
\label{tab:instruction-dimensions-chisq}
\begin{tabular}{@{}lccccccc@{}}
\toprule
Instruction dimension &
Functional (0 / 1) &
Structural (0 / 1) &
Total $N$ &
$\chi^2(1)$ &
$p$ \\
\midrule
Tone / style shift &
154 / 13 &
133 / 15 &
315 &
0.54 &
.464 \\

Qualifications articulation &
51 / 105 &
34 / 102 &
315 &
1.30 &
.254 \\

Structure prompting &
137 / 30 &
123 / 25 &
315 &
0.06 &
.802 \\
\bottomrule
\end{tabular}
\end{table}

Table \ref{tab:instruction-dimensions-binomial} reports binomial tests evaluating whether the distribution of instruction types across conditions deviated from an expected 50/50 split. For each instruction dimension, the table shows the number of instructions coded as present in each condition and the corresponding two-tailed binomial test results.

\begin{table}[!ht]
\centering
\scriptsize
\def\arraystretch{1.1}
\setlength{\tabcolsep}{0.25em}
\caption{Binomial tests evaluating deviations from a 50/50 distribution across conditions for each instruction dimension.}
\label{tab:instruction-dimensions-binomial}
\begin{tabular}{@{}lcccc@{}}
\toprule
Instruction dimension &
Functional ($n$) &
Structural ($n$) &
Total $N$ &
Binomial $p$ \\
\midrule
Tone / style shift (present) &
13 &
15 &
28 &
.851 \\

Qualifications articulation (present) &
112 &
108 &
220 &
.840 \\

Structure prompting (present) &
30 &
25 &
55 &
.590 \\
\bottomrule
\end{tabular}
\end{table}

\section{Grammatical Error Statistical Analysis Details}
\label{gramm-error-appendix}
In this section, we provide more details on the exploratory statistical analysis we conducted to understand why the cover letters written by structural participants had more grammatical, spelling and punctuation errors.

Table \ref{tab:anova-language} reports exploratory two-way analyses of variance examining whether differences in erroneous suggestion acceptance measures persisted after controlling for participants’ native language. For each dependent variable, the table reports the main effects of condition and native language, as well as their interaction.

Table \ref{tab:grammar-errors} reports descriptive statistics and inferential test results for grammatical correctness and erroneous suggestion acceptance measures underlying Figure~\ref{fig:grammar-corr}. The table summarizes group-level means and standard deviations for the functional and structural conditions, along with the statistical tests, test statistics, $p$-values, and effect sizes used to evaluate differences between conditions.

\begin{table}[h]
\centering
\scriptsize
\def\arraystretch{1.1}
\setlength{\tabcolsep}{0.4em}
\caption{Two-way ANOVA results for erroneous suggestion acceptance measures, controlling for native language.}
\label{tab:anova-language}
\begin{tabular}{@{}lcccc@{}}
\toprule
Dependent variable &
Effect &
$F(1,44)$ &
$p$ &
Partial $\eta^2$ \\
\midrule
Error ratio &
Condition (MMNumeric) &
2.36 &
.132 &
.051 \\
 &
Native language &
1.58 &
.216 &
.035 \\
 &
Condition $\times$ Language &
0.41 &
.526 &
.009 \\

\midrule
Errors per word  &
Condition (MMNumeric) &
3.37 &
.073 &
.071 \\
 &
Native language &
1.32 &
.258 &
.029 \\
 &
Condition $\times$ Language &
0.36 &
.554 &
.008 \\
\bottomrule
\end{tabular}
\end{table}

\begin{table*}[tbh!]
\centering
\scriptsize
\def\arraystretch{1.1}
\setlength{\tabcolsep}{0.25em}
\caption{Grammatical correctness and erroneous suggestion acceptance measures by condition.}
\label{tab:grammar-errors}
\begin{tabular}{@{}lcccccc@{}}
\toprule
Measure &
Functional condition &
Structural condition &
Test &
Statistic &
$p$ &
Effect size \\
\midrule
Calibrated corrections &
$M = 0.016$, $SD = 0.010$ &
$M = 0.023$, $SD = 0.010$ &
$t(46)$ &
$t = -2.32$ &
.013 &
$d = 0.67$ \\

Error ratio &
$M = 0.146$, $SD = 0.119$ &
$M = 0.226$, $SD = 0.157$ &
Mann--Whitney $U$ &
$U = 204.5$, $Z = -1.73$ &
.084 &
$r = 0.25$ \\

Errors per word &
$M = 0.006$, $SD = 0.007$ &
$M = 0.011$, $SD = 0.008$ &
Mann--Whitney $U$ &
$U = 170$, $Z = -2.44$ &
.015 &
$r = 0.35$ \\
\bottomrule
\end{tabular}

\end{table*}

\section{Cover Letter Qualitative Quality Assessment Details}
\label{cover-letter-quality-appendix}
This subsection provides additional detail on the qualitative evaluation of cover letter outputs produced during the study.
Table \ref{tab:cover-letter-quality} reports descriptive statistics and nonparametric test results for qualitative  evaluations of cover letter quality by condition. Letters were assessed and rated across three dimensions---relevance, flow and clarity, and tone/style appropriateness---and ratings distributions were then compared between the functional and structural conditions using Mann--Whitney $U$ tests.

\begin{table*}[h!t]
\centering
\scriptsize
\def\arraystretch{1.1}
\setlength{\tabcolsep}{0.25em}
\caption{Cover letter quality ratings by condition.}
\label{tab:cover-letter-quality}
\begin{tabular}{@{}lcccccc@{}}
\toprule
Measure &
Functional condition &
Structural condition &
Test &
Statistic &
$p$ &
Effect size \\
\midrule
Relevance (Q1) &
$M = 3.67$, $SD = 0.87$ &
$M = 3.83$, $SD = 0.96$ &
Mann--Whitney $U$ &
$U = 249$, $Z = -0.85$ &
.398 &
$r = -0.12$ \\

Flow (Q2) &
$M = 4.04$, $SD = 0.81$ &
$M = 3.96$, $SD = 0.91$ &
Mann--Whitney $U$ &
$U = 280.5$, $Z = -0.16$ &
.869 &
$r = -0.02$ \\

Tone / style (Q3) &
$M = 4.17$, $SD = 0.64$ &
$M = 4.04$, $SD = 0.96$ &
Mann--Whitney $U$ &
$U = 285.5$, $Z = -0.06$ &
.954 &
$r = -0.01$ \\
\bottomrule
\end{tabular}
\end{table*}


\section{Post-task Survey Statistical Analysis Details}
\label{post-task-appendix}
This subsection reports the full statistical details for the post-task self-report measures collected after participants completed the writing task. 
Table~\ref{tab:composite-scales} reports the full breakdown of the two composite self-report measures used in the study—\textit{Usefulness} and \textit{Perceived ownership}. For each composite, we list the constituent items, internal reliability, and group-level descriptive statistics by condition, along with the corresponding independent-samples $t$-tests. These results provide transparency into how each composite measure was constructed and confirm that there were no statistically significant differences between conditions at either the scale or item level.

\begin{table*}[ht]
\centering
\scriptsize
\caption{Composite self-report scales and constituent items by condition.}
\label{tab:composite-scales}
\begin{tabular}{@{}llcccc@{}}
\toprule
Scale / Item &
Type &
Functional condition &
Structural condition &
Test statistic &
$p$ (one-sided) \\
\midrule
\multicolumn{6}{l}{\textit{Usefulness} ($\alpha = .76$)} \\

Usefulness (composite) &
Scale mean &
$M = 3.89$, $SD = 0.56$ &
$M = 4.01$, $SD = 0.55$ &
$t(46) = -0.78$ &
.220 \\

Suggestions were relevant &
Item &
$M = 3.88$, $SD = 0.68$ &
$M = 4.04$, $SD = 0.46$ &
$t(46) = -0.99$ &
.163 \\

Suggestions were helpful &
Item &
$M = 3.96$, $SD = 0.69$ &
$M = 4.13$, $SD = 0.68$ &
$t(46) = -0.84$ &
.202 \\

Customizing suggestions using the Instruction Box was helpful &
Item &
$M = 3.75$, $SD = 0.79$ &
$M = 3.79$, $SD = 0.66$ &
$t(46) = -0.20$ &
.422 \\

Able to customize the provided suggestions &
Item &
$M = 3.96$, $SD = 0.96$ &
$M = 4.08$, $SD = 0.83$ &
$t(46) = -0.48$ &
.315 \\

\midrule
\multicolumn{6}{l}{\textit{Perceived ownership} ($\alpha = .71$)} \\

Perceived ownership (composite) &
Scale mean &
$M = 3.95$, $SD = 0.64$ &
$M = 3.93$, $SD = 0.78$ &
$t(46) = 0.10$ &
.460 \\

I am the main contributor to the content of the cover letter &
Item &
$M = 4.29$, $SD = 0.69$ &
$M = 4.08$, $SD = 0.93$ &
$t(46) = 0.88$ &
.191 \\

I am accountable for all aspects of the cover letter &
Item &
$M = 3.63$, $SD = 1.13$ &
$M = 3.67$, $SD = 1.17$ &
$t(46) = -0.13$ &
.450 \\

I feel like I am the author of the cover letter &
Item &
$M = 3.88$, $SD = 0.85$ &
$M = 4.04$, $SD = 0.96$ &
$t(46) = -0.64$ &
.263 \\

I feel like the cover letter represents me &
Item &
$M = 4.00$, $SD = 0.89$ &
$M = 3.92$, $SD = 1.06$ &
$t(46) = 0.30$ &
.384 \\

\bottomrule
\end{tabular}
\end{table*}

Table~\ref{tab:single-item-selfreport} presents descriptive and inferential statistics for all remaining single-item self-report measures that were not included in composite scales. These items capture participants’ perceptions of system usability, suggestion quality, perceived control, and authorship. Consistent with the results reported in the main paper, none of these measures showed statistically significant differences between the functional and structural conditions.

\begin{table*}[t]
\centering
\scriptsize
\caption{Single-item self-report measures by condition.}
\label{tab:single-item-selfreport}
\begin{tabular}{@{}lcccc@{}}
\toprule
Measure &
Functional condition &
Structural condition &
$t(46)$ &
$p$ (one-sided) \\
\midrule
CoAuthor was easy to use &
$M = 3.88$, $SD = 1.04$ &
$M = 4.38$, $SD = 0.58$ &
$-2.07$ &
\textbf{.022} \\

The final letter is error-free &
$M = 3.38$, $SD = 1.06$ &
$M = 3.46$, $SD = 1.22$ &
$-0.25$ &
.400 \\

The final letter has appropriate style and content &
$M = 3.96$, $SD = 1.00$ &
$M = 4.25$, $SD = 0.79$ &
$-1.12$ &
.134 \\

The suggestions influenced the content of the cover letter &
$M = 3.17$, $SD = 1.17$ &
$M = 3.50$, $SD = 0.93$ &
$-1.09$ &
.140 \\

I felt in control of the quality and content of the suggestions &
$M = 3.25$, $SD = 1.07$ &
$M = 2.92$, $SD = 1.18$ &
$1.03$ &
.155 \\

I felt in control of what was included in the final letter &
$M = 4.38$, $SD = 0.92$ &
$M = 4.46$, $SD = 0.66$ &
$-0.36$ &
.360 \\

I felt like I was writing the text and CoAuthor was assisting me &
$M = 3.79$, $SD = 0.66$ &
$M = 4.00$, $SD = 1.06$ &
$-0.82$ &
.209 \\

I felt like CoAuthor was writing the text and I was assisting &
$M = 1.96$, $SD = 1.00$ &
$M = 2.08$, $SD = 1.02$ &
$-0.43$ &
.335 \\

The suggestions were error-free &
$M = 1.71$, $SD = 0.75$ &
$M = 2.13$, $SD = 1.15$ &
$-1.48$ &
.072 \\

The suggestions were helpful &
$M = 3.96$, $SD = 0.69$ &
$M = 4.13$, $SD = 0.68$ &
$-0.84$ &
.202 \\

\bottomrule
\end{tabular}
\end{table*}

\end{document}